\documentclass[12pt]{article}
\usepackage{amsmath}
\usepackage{graphicx}
\usepackage{enumerate}
\usepackage{natbib}
\usepackage{gensymb} 
\usepackage{url} 
\usepackage[margin=1in]{geometry}
\usepackage{needspace}
\usepackage{mathtools}
\usepackage{empheq}

\newcommand{\blind}{1}

\usepackage{amssymb}
\usepackage{subcaption}
\usepackage{xcolor}

 \newcommand{\overbar}[1]{\mkern 1.5mu\overline{\mkern-1.5mu#1\mkern-1.5mu}\mkern 1.5mu}
\newcommand{\bbR}{\mathbb{R}}
\newcommand{\bbZ}{\mathbb{Z}}

\newcommand{\bA}{\mathbf{A}}
\newcommand{\bH}{\mathbf{H}}
\newcommand{\bD}{\mathbf{D}}
\newcommand{\bI}{\mathbf{I}}
\newcommand{\bX}{\boldsymbol{X}}
\newcommand{\bY}{\boldsymbol{Y}}
\newcommand{\bZ}{\boldsymbol{Z}}

\newcommand{\bT}{\boldsymbol{T}}
\newcommand{\bC}{\mathbf{C}}
\newcommand{\bx}{\boldsymbol{x}}
\newcommand{\by}{\boldsymbol{y}}

\newcommand{\bs}{\boldsymbol{s}}
\newcommand{\bS}{\mathbf{S}}
\newcommand{\ba}{\boldsymbol{a}}
\newcommand{\bB}{\boldsymbol{B}}
\newcommand{\bb}{\boldsymbol{b}}
\newcommand{\bu}{\boldsymbol{u}}
\newcommand{\bv}{\boldsymbol{v}}
\newcommand{\bom}{\boldsymbol{m}}

\newcommand{\bV}{\boldsymbol{V}}
\newcommand{\bt}{\boldsymbol{t}}

\newcommand{\bmu}{\boldsymbol{\mu}}

\newcommand{\bh}{\boldsymbol{h}}

\newcommand{\btheta}{\boldsymbol{\theta}}
\newcommand{\bLambda}{\boldsymbol{\Lambda}}
\newcommand{\bSigma}{\boldsymbol{\Sigma}}

\newtheorem{Assumption}{Assumption}
\usepackage[ruled,vlined]{algorithm2e}

\begin{document}

\def\spacingset#1{\renewcommand{\baselinestretch}%
{#1}\small\normalsize} \spacingset{1}


\if1\blind
{
  \title{\bf Are You All Normal? It Depends!}
 \author{\hspace{-1.5cm}Wanfang Chen\thanks{Wanfang Chen is an Assistant Professor at the Academy of Statistics and Interdisciplinary Sciences, East China Normal University, Shanghai, China (e-mail: wfchen@fem.ecnu.edu.cn); and Marc G. Genton is a Distinguished Professor of Statistics at the Statistics Program, KAUST, Thuwal, Saudi Arabia (e-mail: marc.genton@kaust.edu.sa). This research was supported by the National Key Research and Development Program (2021YFA1000101), Zhejiang Provincial Natural Science Foundation of China (LZJWY22E090009), Natural Science Foundation of Shanghai (22ZR1420500), the Open Research Fund of Key Laboratory of Advanced Theory and Application in Statistics and Data Science-MOE, ECNU and KAUST.}  \, and \,Marc G. Genton$^{**}$\\
\hspace{-1.15cm}$^{*}$Academy of Statistics and Interdisciplinary Sciences, East China Normal University\\
\hspace{-1.15cm}$^{**}$Statistics Program, King Abdullah University of Science and Technology}
  \maketitle
} \fi

\if0\blind
{
  \bigskip
  \bigskip
  \bigskip
  \begin{center}
    {\LARGE\bf Are You All Normal? It Depends!}
\end{center}
  \medskip
} \fi

\vspace{-1cm}
\spacingset{1.5}
\begin{abstract}
The assumption of normality has underlain much of the development of statistics, including spatial statistics, and many tests have been proposed. In this work, we focus on the multivariate setting and first review the recent advances in multivariate normality tests for i.i.d. data, with emphasis on the skewness and kurtosis approaches. We show through simulation studies that some of these tests cannot be used directly for testing normality of spatial data. 
We further review briefly the few existing univariate tests under dependence (time or space), and then propose a new multivariate normality test for spatial data by accounting for the spatial dependence. The new test utilizes the union-intersection principle to decompose the null hypothesis into intersections of univariate normality hypotheses for projection data, and it rejects the multivariate normality if any individual hypothesis is rejected. The individual hypotheses for univariate normality are conducted using a Jarque-Bera type test statistic that accounts for the spatial dependence in the data. We also show in simulation studies that the new test has a good control of the type I error and a high empirical power, especially for large sample sizes. We further illustrate our test on bivariate wind data over the Arabian Peninsula.
\end{abstract}

\noindent%
\vfill

\newpage
\spacingset{1.5} 
\section{Introduction}\label{sec:intro}
Normality is one of the most commonly made assumptions in the development and use of statistical procedures, such as t-tests, tests for regression coefficients, the F-test of homogeneity of variance, discriminant analysis and analysis of variance (ANOVA). The performance of these procedures can be affected to various extents if the normality assumption is violated (see, e.g., \cite{pitman1938significance}, \cite{geary1947testing}, \cite{box1953non}, \cite{tukey1960survey}, \cite{subrahmaniam1975robustness}, \cite{d1977robustness}, and \cite{looney1995use}). Hence, the problem of testing whether a sample of observations comes from a normal distribution or not has received much attention, and numerous methods for testing for normality have been developed. There is now a very large body of literature on tests for univariate normality; for a review of classical tests, see, e.g., \cite{mardia19809}, \cite{d1986goodness} and \cite{thode2002testing}, and for comparative studies on the power of selected normality tests, see, e.g., \cite{shapiro1968comparative}, \cite{pearson1977tests}, \cite{keskin2006comparison}, \cite{oztuna2006investigation}, \cite{farrell2006comprehensive}, \cite{thadewald2007jarque}, \cite{yazici2007comparison}, \cite{romao2010empirical}, \cite{yap2011comparisons}, \cite{noughabi2011monte}, \cite{ahmad2015power}, \cite{islam2017stringency} and \cite{sanchez2019graphical}. 

Relatively less work has been done in the field of testing for multivariate normality (MVN) compared to that done for the univariate case, since there can be many difficult cases for MVN; for instance, non-normal distributions that have all lower-dimensional marginals being normal (see, e.g., \cite{dutta2014non}). In addition, classical univariate normality tests, such as the $\chi^2$-test, have limited applicability in higher dimensions. Reviews on the tests for MVN have been given by \cite{thode2002testing}, \cite{henze2002invariant} and \cite{ebner2020tests}, with the last one emphasizing on several classes of the weighted $L^2$-statistics. Evaluation on the power of various tests for MVN is quite sparse, and among the more comprehensive studies are those of \cite{horswell1992comparison}, \cite{romeu1993comparative}, \cite{mecklin2005monte}, \cite{farrell2007tests}, \cite{joenssen2014power} and \cite{hanusz2018monte}. The Jarque-Bera (JB) type test \citep{bera1981efficient}, which combines the sample skewness and kurtosis measures, is among one of the most commonly used tests due to its simplicity and good power properties.

In spatial statistics applications, the Gaussian assumption is also widely used to improve finite-sample inference and effectively employ Bayesian methods. \cite{zimmerman2010classical} and \cite{gelfand2016spatial} provided surveys of Gaussian modeling in spatial statistics. Recent research has focused on applying spatial statistical methods based on the Gaussian assumption to large datasets and advancing computational approaches; see, e.g., \cite{nychka2015multiresolution}, \cite{paciorek2015parallelizing}, \cite{katzfuss2017multi} and \cite{guhaniyogi2018meta}. 
Despite the prevalence of the Gaussian assumption made in spatial statistics, there appears to be very few significance tests that could be used to assess if it is reasonable to assume that a given spatial dataset can be treated as a realization of a Gaussian random field. 
All the aforementioned tests cannot be directly used for spatial data, since they are designed for examining the normality in a random sample (i.e., i.i.d. observations), so that the conventional large-sample approximations to the null distributions of the test statistics are either unknown or inaccurate under spatial dependence. In this work, we show by simulation studies in Section~\ref{sec:simu} that the sample skewness and kurtosis deviate from their theoretical values in the i.i.d. case as the degree of spatial dependence increases. Hence, the usual test of normality based on the sample skewness and kurtosis may be misleading if the observations in the sample are dependent, as also indicated by the severely inflated type I error from our simulation study in Section~\ref{sec:newtest}. 

A review on univariate normality tests for data with serial dependence in time series is given by \cite{psaradakis2020normality}, but these tests need to be justified, extended or modified if they are to be applied to spatial data, and further generalized to the multivariate setting, which is not always possible. \cite{pardo2004normality} demonstrated a methodology for the application of standard univariate normality tests, such as the Kolmogorov--Smirnov test, the chi-square test, and the Shapiro--Wilks test, to spatially correlated data, using block kriging in de-clustering to obtain unbiased estimates of the probability density function or the cumulative density function. \cite{olea2009kolmogorov} and \cite{zheng2019kolmogorov} investigated the Kolmogorov-Smirnov test under spatial correlations, using bootstrap methods or Monte Carlo procedures. However, these tests are either difficult to implement or computationally intensive. \cite{horvath2020testing} developed a JB-type test for spatial data defined on a grid under the assumption of stationarity by accounting for the spatial dependence of the observations. The test is easy to implement, shown to have good empirical size and power, and can be justified asymptotically. To our knowledge, no normality test for multivariate spatial data has been proposed yet.

The goal of this study is twofold. First, we aim at providing a comprehensive review on recent MVN tests for i.i.d. data based on skewness and kurtosis approaches, proposed since the review works by \cite{thode2002testing} and \cite{henze2002invariant}. Second, we propose a MVN test for spatially correlated data by extending the test of \cite{horvath2020testing} to the multivariate setting. We consider the practically common case where the data to be tested are the zero-mean residuals of regression and spatial models. The type I error and empirical power of the new test are assessed by simulation studies. In the title of our paper, the ``All'' represents ``multivariate'', and the answer to the question of testing multivariate normality ``depends'' on the underlying dependence (in space or time). 

The rest of this paper is organized as follows. Section~\ref{sec:pre} introduces some useful preliminaries, terminologies and notations. Section~\ref{sec:review} reviews the recent developments of MVN tests based on the skewness and kurtosis approaches in the i.i.d. setting; Section~\ref{sec:other} reviews the chi-squared type and BHEP-type tests, and the other types of MVN test for i.i.d. data are presented in the Supplementary Material. Section~\ref{sec:simu} demonstrates a simulation study to investigate the influence of spatial dependence on the measures of skewness and kurtosis for multivariate Gaussian random fields. Section~\ref{sec:newtest} describes our new test for MVN under spatial dependence and its performance based on the type I error and empirical power. Section~\ref{sec:app} describes a data application based on bivariate wind data over the Arabian Peninsula. Section~\ref{sec:disc} concludes and discusses future work directions.

\section{Preliminaries, Terminologies and Notations}\label{sec:pre}
In this section, we describe the preliminaries, terminologies and notations that will be used throughout this paper.

The significance testing problem is formulated as follows. Let $\bX_i\in\bbR^{p},i=1,\ldots,n$, be observations (a random sample or spatially correlated data) from a $p$-variate distribution with cumulative distribution function (CDF) $F_{\bX}$. Let $\mathcal{N}_{p}(\boldsymbol{\mu},\bSigma)$ denote the $p$-variate normal distribution with expectation $\bmu$ and nonsingular covariance matrix $\bSigma$, and let $\mathcal{N}_p$ denote the class of all non-degenerate $p$-variate normal distributions. Our interest is to test, based on the observations $\bX_1,\ldots, \bX_n$, the hypothesis $H_0: F_{\bX}\in \mathcal{N}_p$, against general alternatives.

It is usually desired that the tests for MVN possess the properties of affine invariance and universal consistency. Since the class $\mathcal{N}_p$ is closed with respect to full rank affine transformations, in order to ensure the same conclusion regarding rejection or non-rejection of $H_0$ given the original data $\bX_1,\ldots, \bX_n$ and the transformed data $\bA\bX_1 + \bb,\ldots, \bA\bX_n + \bb$, where $\bA\in\bbR^{p\times p}$ is nonsingular and $\bb\in\bbR^{p}$, any test statistic $T_n(\bX_1,\ldots,\bX_n)$ should be affine invariant, i.e., $T_n(\bA\bX_1+\bb,\ldots,\bA\bX_n+\bb)=T_n(\bX_1,\ldots,\bX_n)$.
The consistency class of a test statistic $T_n$ for $H_0$ is the set of probability distributions $P$ over $\bbR^{p}$ such that, if the underlying distribution is $P$, the probability of rejecting $H_0$ tends to one as the sample size $n$ goes to infinity, when using the test statistic $T_n$. As the alternatives to normality are rarely known in practice, it is important that the consistency class of a test for MVN is the set of all $P\notin \mathcal{N}_p$, which implies that the test is able to detect any non-normal alternative distribution, at least for large samples. Here, we call a test to be universally consistent if it is consistent against any fixed non-normal alternative distributions.

Since there are, in principle, an infinite number of alternatives to normal distributions, no uniformly most powerful test exists for MVN. Therefore, two types of tests are developed tailored to the problem of interest. One type consists of {\it omnibus} tests that are designed to cover all possible alternatives, usually with only reasonably high and generally suboptimal powers. Most of the tests in the literature are omnibus tests. The other type refers to {\it directed} tests that are highly powerful for some specific classes of alternatives, at the cost of being blind to other types of alternatives. Combinations of directed tests have also been suggested as omnibus tests. Tests based on measures of multivariate skewness and kurtosis are typically directed tests, and they have certain diagnostic limitations as clarified by \cite{henze2002invariant} and also mentioned in Section~\ref{sec:review}. Nevertheless, one important role of directed tests is that they can be used to detect types of departures from normality that are most dangerous in the underlying problem. For example, the size of the Hotelling $T^2$ test \citep{hotelling1931generalization} is much influenced by the asymmetry of the distribution, while symmetric departures from normality are not so crucial \citep{mardia1970measures}. In addition, for restricted families of alternatives that are closed under the action of some groups of transformations, it may be possible to construct most powerful invariant (MPI) tests and thus set benchmarks for assessing the performance of other invariant tests.

In what follows, let $\textbf{\textit{0}}$ denote the null vector of length $p$, $\bI_p$ denote the identity matrix of size $p\times p$, $\|\cdot\|$ denote the Euclidean norm in $\bbR^{p}$, and a superscript $\top$ denote a transpose. Also, denote the sample mean vector and sample covariance matrix, for the $p$-variate observations $\bX_1,\ldots,\bX_n$, as $\overbar{\bX}=\frac{1}{n}\sum_{i=1}^{n}\bX_i$ and $\bS=\frac{1}{n}\sum_{i=1}^{n}(\bX_{i}-\overbar{\bX})(\bX_{i}-\overbar{\bX})^{\top}$, respectively, and $\widetilde{\bS}=\frac{n}{n-1}\bS$ is the unbiased sample covariance matrix. In addition, assume that $n\geq p+1$ so that $\bS$ is invertible with probability one \citep{eaton1973non}. Denote by $\bS^{-1/2}$ the unique symmetric square root of $\bS$, and define the scaled residuals as $\bY_{i}=\bS^{-1/2}(\bX_{i}-\overbar{\bX}),i=1,\ldots,n$, which are asymptotically $\mathcal{N}_{p}(\textbf{\textit{0}},\bI_{p})$ under $H_0$.

\section{Recent Advances of MVN Tests Based on Skewness and Kurtosis Approaches for I.I.D. Data} \label{sec:review}
Recent work on MVN tests for i.i.d. data can be classified into five categories: 1) skewness and kurtosis approaches, 2) chi-squared type tests, 3) BHEP-type tests based on the empirical characteristic function, 4) other generalizations of univariate normality tests, and 5) multiple testing procedures that combine multiple tests for MVN. In this section, we review the first category, i.e., tests based on skewness and kurtosis measures. In the next section, we review the chi-squared type and BHEP-type tests, and also present the review for the remaining two categories in the Supplementary Material for readers' reference. 

In univariate statistics, the skewness and kurtosis of a random variable $X$, with mean $\mu$ and variance $\sigma^2$, are defined as
\begin{align*}
\beta_1&=\mathbb{E}\left\{ \left(\frac{X-\mu}{\sigma} \right)^3 \right\}=\frac{\mu_{3}}{\mu_{2}^{3/2}}, \mbox{  and } \beta_2=\mathbb{E}\left\{ \left(\frac{X-\mu}{\sigma} \right)^4 \right\}=\frac{\mu_{4}}{\mu_{2}^{2}},
\end{align*}
respectively, where $\mu_{i}$ is the $i$th central moment of $X$. For a normal distribution, $\beta_1=0$ and $\beta_2=3$. Hence, $\beta_{2}-3$ is called excess kurtosis with respect to a normal distribution. The skewness $\beta_1=0$ for symmetric distributions and $\beta_1>0$ $(<0)$ for right (left)-asymmetric distributions, while the kurtosis $\beta_2=3$ for the normal distribution, and $\beta_2>3$ $(<3)$ for distributions that are heavier-tailed (lighter-tailed) than the normal one.

Tests based on the univariate sample skewness and kurtosis are among the earliest procedures for assessing univariate normality. Due to their popularity and good power properties, some of the first tests for MVN are based on extensions of the notion of skewness and kurtosis to the multivariate setting. The Mardia's tests \citep{mardia1970measures, mardia1974applications} are perhaps the most often referenced tests for MVN. \cite{mardia1970measures} firstly extended the measures of skewness and kurtosis of a $p$-dimensional random vector $\boldsymbol{X}=(X_1,X_2,\ldots,X_p)^{\top}$, with mean vector $\boldsymbol{\mu}$ and covariance matrix $\boldsymbol{\Sigma}$, as
\begin{align*}
\beta_{1,p}  = \mathbb{E}\left[\left\{(\boldsymbol{X}-\boldsymbol{\mu})^{\top}\boldsymbol{\Sigma}^{-1}(\boldsymbol{Y}-\boldsymbol{\mu}) \right\}^3  \right] \;\; \mbox{ and }\;\;
\beta_{2,p} = \mathbb{E}\left[\left\{(\boldsymbol{X}-\boldsymbol{\mu})^{\top}\boldsymbol{\Sigma}^{-1}(\boldsymbol{X}-\boldsymbol{\mu}) \right\}^2  \right], 
\end{align*}
respectively, where $\boldsymbol{X}$ and $\boldsymbol{Y}$ are independently and identically distributed random vectors. For a $p$-variate normal distribution, $\beta_{1,p}=\boldmath{0}$ and $\beta_{2,p}=p(p+2)$. For all distributions, $\beta_{1,p}\geq 0$, and for $p=1$, $\beta_{1,p}$ reduces to the square of the univariate skewness. The sample measures are also defined for i.i.d. samples, $\boldsymbol{X}_{i},\,i=1,\ldots,n$, as
\begin{align*}
b_{1,p} &= \frac{1}{n^2}\sum_{i=1}^{n}\sum_{j=1}^{n}\left\{(\boldsymbol{X}_{i}-\bar{\boldsymbol{X}})^{\top}\bS^{-1}(\boldsymbol{X}_{j}-\bar{\boldsymbol{X}}) \right\}^3=\frac{1}{n^2}\sum_{i=1}^{n}\sum_{j=1}^{n}(\bY_{i}^{\top}\bY_j)^{3},\\
b_{2,p} &= \frac{1}{n}\sum_{i=1}^{n}\left\{(\boldsymbol{X}_{i}-\bar{\boldsymbol{X}})^{\top}\bS^{-1}(\boldsymbol{X}_{i}-\bar{\boldsymbol{X}}) \right\}^2=\frac{1}{n}\sum_{i=1}^{n}(\bY_{i}^{\top}\bY_{i})^2.
\end{align*}
 \cite{mardia1970measures} then proposed tests based on $b_{1,p}$ and $b_{2,p}$ as:
\begin{equation}
\textrm{MS}=nb_{1,p}/6, \;\;\textrm{MK}=\{b_{2,p}-p(p+2)\}/\{8p(p+2)/n\}^{1/2}, 
\label{eq:MS}\\
\end{equation}
which are asymptotically $\chi_{p(p+1)(p+2)/6}^2$ and $\mathcal{N}(0,1)$, respectively, under $H_0$. Other classical measures of multivariate skewness and kurtosis and related tests for MVN have been proposed by, for example, \cite{malkovich1973tests}, \cite{isogai1982measure}, \cite{srivastava1984measure}, \cite{koziol1987alternative} and \cite{mori1994multivariate}. 

Univariate normality tests often use classical measures of asymmetry based on the standardized distance between two separate location parameters, and measures of kurtosis based on the ratios of two scale measures, such as the classical standardized fourth moment. Motivated by these facts, \cite{kankainen2007tests} proposed a measure of multivariate skewness based on the Mahalanobis distance between two multivariate location vector estimates, and a measure of multivariate kurtosis based on the (matrix) distance between two scatter matrix estimates. A vector-valued (matrix-valued) statistic is called a location vector (a scatter matrix) if it is affine equivariant (see Section 2 in \cite{kankainen2007tests}). Then, the test statistic for MVN (to detect skewness) is given by $U=(\bT_1-\bT_2)^{\top}\bC^{-1}(\bT_1-\bT_2)$,
where $\bT_1$ and $\bT_2$ are two separate location vectors and $\bC$ is a scatter matrix, and the kurtosis test statistic is given by
\begin{align*}
W&=\|\bC_{1}^{-1}\bC_2-\bI_{p} \|^2=\big[ \mbox{tr}\{(\bC_{1}^{-1}\bC_2)^2\}-\frac{1}{p}\mbox{tr}^2(\bC_{1}^{-1}\bC_2)\big]+\frac{1}{p}\big\{\mbox{tr}(\bC_{1}^{-1}\bC_2)-p\big\}^2,
\end{align*}
where $\|\cdot\|^2=\mbox{tr}(\cdot^{\top}\cdot)$, and $\bC_1$ and $\bC_2$ are two separate scatter matrices.   
Using special choices of location and scatter estimators, it is possible to obtain generalizations of classical Mardia's measures of multivariate skewness and kurtosis. 

\cite{thulin2014tests} proposed a measure of multivariate skewness in a way that resembles the construction in \cite{mardia1970measures}. 
For the sample $\bX_1,\ldots, \bX_n$, write $\overbar{\bX}=(\overbar{X}_1,\ldots,\overbar{X}_p)^{\top}$, $\bS=\{S_{ij}\}$, and $\bu=(S_{11},\ldots,S_{pp},S_{12},\ldots,S_{1p},\ldots,S_{2p},\ldots,S_{p-1,p})^{\top}$. It is well known that $\overbar{\bX}$ and $\bu$ are independent under $H_0$. Denote the covariance matrix of $\overbar{\bX}$ and $\bu$ by $\mbox{Cov}(\overbar{\bX},\bu)=\left[\begin{matrix}
\bLambda_{11} & \bLambda_{12} \\
\bLambda_{21} & \bLambda_{22}
\end{matrix}\right]$,
where $\bLambda_{11}$ is the covariance matrix of $\overbar{\bX}$ and so on.
The canonical correlations, $\lambda_1,\ldots,\lambda_p$, of $\overbar{\bX}$ and $\bu$ are the square roots of the eigenvalues of $\bLambda_{11}^{-1}\bLambda_{12}\bLambda_{22}^{-1}\bLambda_{21}$, and they are all equal to zero under $H_0$. The measure of multivariate skewness proposed by \cite{mardia1970measures} is based on the sum of squared canonical correlations:
\begin{align}
\beta_{1,p}=2\sum_{i=1}^{p}\lambda_{i}^2=2\;\mbox{tr}(\bLambda_{11}^{-1}\bLambda_{12}\bLambda_{22}^{-1}\bLambda_{21}),
\label{eq:b1p}
\end{align}
under the assumption that the cumulants of order higher than $3$ of $\bX$ are negligible. The sample counterpart of $\beta_{1,p}$ can be used to construct tests for MVN. 
\cite{thulin2014tests} derived explicit expressions for the elements of $\mbox{Cov}(\overbar{\bX},\bu)$ in terms of the moments of $(X_1,\ldots,X_p)$ (see his Theorem 1), and proposed a new test, $Z_{2,p}^{HL}$, based on the sample counterpart of $\mbox{Cov}(\overbar{\bX},\bu)$ (see his Equation (12)). The author constructed another test based on the fact that $\overbar{\bX}$ and $\bv=(S_{111},S_{112},\ldots,S_{p,p,(p-1)},S_{ppp})^{\top}$ are also independent under $H_0$, where 
\[
S_{ijk}=\frac{n}{(n-1)(n-2)}\sum_{r=1}^{n}(X_{r,i}-\overbar{X}_i)(X_{r,j}-\overbar{X}_j)(X_{r,k}-\overbar{X}_k).
\] 
\cite{yamada2015kurtosis} generalized Mardia's multivariate kurtosis for testing MVN when the data consist of a random sample of two-step monotone incomplete observations.

One disadvantage of the above tests is that they only consider departures from multivariate normality revealed by skewness and kurtosis, and failure to reject the null hypothesis leaves open the question of whether there are departures from normality in other ways. Consequently, these tests are not universally consistent. For example, the test based on multivariate kurtosis in the sense of \cite{malkovich1973tests} is inconsistent against spherically symmetric alternative distributions with normal marginal kurtosis, $3$. Furthermore, these tests rely only on asymptotic properties, that is, they require large samples to achieve both reasonably accurate control of type I error and high power.

The omnibus Jarque-Bera (JB)-type tests address the above issue by combining the skewness and kurtosis measures. The univariate JB test \citep{bera1981efficient}, based on a univariate random sample $X_i\in\bbR,i=1,\ldots,n$, is given by $\textrm{JB}=\frac{nb_1^2}{6}+\frac{n(b_2-3)^2}{24}$,
where $b_1$ and $b_2$ are the sample skewness and kurtosis, respectively, given by $b_{1}=\frac{\sqrt{n(n-1)}}{n-2}\frac{m_3}{m_2^{3/2}}$ and $b_2=\frac{m_4}{m_2^{2}}$, where $m_k=\frac{1}{n}\sum_{i=1}^{n}\left(X_i-\frac{1}{n}\sum_{j=1}^{n}X_j \right)^k$. 
Under univariate normality, $\textrm{JB}$ is asymptotically $\chi_{2}^2$.  The simplest way to construct multivariate JB-type tests, based on the sample $\bX_1,\ldots, \bX_n$, is to aggregate individual (univariate) skewness and kurtosis as $\textrm{JB}_\textrm{M}=\sum_{i=1}^{p}\frac{nb_{1(i)}^{2}}{6}+\sum_{i=1}^{p}\frac{n(b_{2(i)}-3)^2}{24}$,
where $b_{1(i)}$ and $b_{2(i)}$ denote the sample skewness and kurtosis, respectively, of component $i$. $\textrm{JB}_\textrm{M}$ is asymptotically distributed as $\chi_{2p}^{2}$ under $H_0$ (see, e.g., \cite{lutkepohl2005new}). However, for both $\textrm{JB}$ and $\textrm{JB}_\textrm{M}$, the sample skewness and kurtosis are not independent in finite samples, and using the asymptotic distribution leads to under-rejection. To remedy this problem, \cite{doornik2008omnibus} proposed to use transformed skewness and kurtosis, which creates statistics that are much closer to standard normal, based on the work of \cite{bowman1975omnibus}. Specifically, the test statistic is
\begin{equation}
\mbox{JB}_{\mbox{\tiny DH}}=\bB_1^{\top}\bB_1+\bB_2^{\top}\bB_2,
\label{eq:JB_DH}
\end{equation}
where $\bB_1=(b_{1(1)}^*,\ldots,b_{1(p)}^*)^{\top}$ and $\bB_2=(b_{2(1)}^*,\ldots,b_{2(p)}^*)^{\top}$ are the transformed vectors of skewness and kurtosis, respectively. 
$\mbox{JB}_{\mbox{\tiny DH}}$ is asymptotically $\chi^{2}_{2p}$ under $H_0$.
\cite{jonsson2011robust} further noticed that there is a pattern of downward size distortions to the test based on $\textrm{JB}_\textrm{M}$; see his Figure 1. He suggested using the test statistic that pools the individual $p$-values:  $\widetilde{\textrm{LM}}=-2\sum_{i=1}^{p}\mbox{ln}(\pi_i)$,
where $\pi_i$ is the $p$-value of the univariate $\textrm{JB}$ test for the $i$th component. $\widetilde{\textrm{LM}}$ has an asymptotic $\chi_{2p}^{2}$ distribution under $H_0$, and simulation studies showed that the previous poor size properties are eliminated (see his Figure 2) without loss of power. The calculation of $\widetilde{\textrm{LM}}$ is somewhat more convenient than using the transformation approach proposed by \cite{doornik2008omnibus}. \cite{kim2016robustified} proposed to aggregate the univariate JB-type statistics based on transformed data.  Suppose the random sample $\bX_1,\ldots,\bX_n$ is from $\mathcal{N}_{p}(\bmu,\bSigma)$. Then the standardized data, $\bZ_i=\bS^{*\top}(\bX_{i}-\overbar{\bX}),\;i=1,\ldots,n$,
follow a $\mathcal{N}_{p}(\textbf{\textit{0}},\bI_p)$ asymptotically under $H_0$, where $\bS^{*}$ is defined by $\bS^{*\top}\bS\bS^{*}=\bI_{p}$. The multivariate test statistics are then formed by adding up the univariate JB-type statistics for each coordinate of the transformed vectors.

Another way to construct multivariate JB-type tests is to combine multivariate skewness and kurtosis measures (see, e.g., \cite{mardia1983omnibus}, \cite{bera1983tests} and \cite{mardia1991rao}). \cite{koizumi2009jarque} proposed two JB-type tests based on multivariate sample skewness and kurtosis of \cite{srivastava1984measure}. For the sample $\bX_1,\ldots, \bX_n$, let $\bS=\bH \bD_{\omega}\bH^{\top}$, where $\bH=(\bh_1 \ldots \bh_p)$ is an orthogonal matrix and $\bD_{\omega}=\mbox{diag}(\omega_1,\ldots,\omega_p)$. The sample measures of multivariate skewness and kurtosis given by \cite{srivastava1984measure} are: 
\begin{align}
 \widetilde{b}_{1,p} = \frac{1}{p}\sum_{i=1}^{p}\left(\frac{m_{3i}}{m_{2i}^{3/2}}\right)^2, \;\;\; \widetilde{b}_{2,p} &= \frac{1}{p}\sum_{i=1}^{p}\frac{m_{4i}}{m_{2i}^{2}}, 
 \label{eq:b1b2}
 \end{align}
respectively, where $m_{ki}=\frac{1}{n}\sum_{j=1}^{n}(Y_{ij}-\overbar{Y}_{i})^{k}$, with $Y_{ij}=\bh_{i}^{\top}\bX_j$ and $\overbar{Y}_{i}=\frac{1}{n}\sum_{j=1}^{n}Y_{ij}$, $i=1,\ldots,p$, $j=1,\ldots,n$. The two JB-type statistics based on $\widetilde{b}_{1,p}$ and $\widetilde{b}_{2,p}$ are:
\begin{align*}
M_1=np\left\{ \frac{\widetilde{b}_{1,p}}{6} + \frac{(\widetilde{b}_{2,p}-3)^2}{24} \right\}\;\;\mbox{ and }\;\;
M_2= \frac{p\widetilde{b}_{1,p}}{\mbox{E}(\widetilde{b}_{1,p})} + \frac{\{\widetilde{b}_{2,p}-\mbox{E}(\widetilde{b}_{2,p})\}^2}{\mbox{Var}(\widetilde{b}_{2,p})},
\end{align*}
both asymptotically $\chi_{p+1}^{2}$  under $H_0$, with $\mbox{E}(\widetilde{b}_{1,p})=\frac{6(n-2)}{(n+1)(n+3)}$, $\mbox{E}(\widetilde{b}_{2,p})=\frac{3(n-1)}{n+1}$, and $\mbox{Var}(\widetilde{b}_{2,p})=\frac{24n(n-2)(n-3)}{p(n+1)^2 (n+3)(n+5)}$ under $H_0$. \cite{enomoto2012multivariate} noticed a difference between the upper percentiles of the distributions of $M_2$ and the $\chi^2$ distribution for small $n$. To mitigate the difference, they proposed a new test statistic by using the variance of $M_2$:
\begin{equation*}
M_3=cM_2 + (1-c)(p+1),
\end{equation*}
which is also asymptotically $\chi_{p+1}^{2}$ under $H_0$, with $c=\left\{\frac{2p(p+1)}{\mbox{Var}(M_2)}\right\}^2$, and $\mbox{Var}(M_2)$ is derived as their Equation (3.1). \cite{koizumi2014modified} suggested two other improved tests of $M_1$ and $M_2$. First, they noticed that in $M_1$, the skewness term asymptotically dominates the kurtosis term for large $p$, so that the omnibus test becomes a directional test for the skewness only. Therefore, they proposed the following test statistic:
\[
\textrm{MJB}_2=z_{WH}^2+\frac{np}{24}(\widetilde{b}_{2,p}-3)^2,
\]
where $z_{WH}=\frac{(z_1/p)^{1/3}-1+2/(9p)}{\sqrt{2/(9p)}}$ is the Wilson-Hilferty transform \citep{wilson1931distribution} of $z_1=np\widetilde{b}_{1,p}/6$. 
When both $p$ and $n$ go to infinity, $\textrm{MJB}_2$ is asymptotically $\chi_{2}^2$ under $H_0$, which does not depend on the dimensionality $p$, and hence the omnibus property of the test is maintained even for large $p$. However, their simulation study showed that the $\textrm{MJB}_2$ test has poor performance in terms of type I error. They further improved $\textrm{MJB}_2$ by a normalizing transform of the sample kurtosis as suggested in \cite{seo2011distribution}:
\[
\textrm{mMJB}=z_{WH}^2+z_{NT}^{2},
\]
where $z_{NT}=\sqrt{\frac{np}{24}}\{-e^{-({\widetilde{b}_{2,p}-3)}}+1+\frac{6}{n}+\frac{12}{np} \}$. The statistic $\textrm{mMJB}$ is asymptotically $\chi_{2}^2$ under $H_0$, and proved to have a more stable behavior in small samples. They further studied the $F$-approximation for $\textrm{mMJB}$ which is shown to be better than the $\chi^2$ approximation, and therefore can be recommended for testing MVN in both small and large samples.

\section{Review of Other Recent MVN Tests For I.I.D. Data}\label{sec:other}
In this section, we review the chi-squared and BHEP-type tests. The remaining types of MVN tests for i.i.d. data (i.e., other generalizations of univariate normality tests and multiple testing procedures that combine multiple tests for MVN) are presented in the Supplementary Material. We summarize some important properties (affine invariance, universal consistency, explicit null distribution) for all the reviewed tests in Table~\ref{table1} at the end of this section.

\subsection{Chi-squared type tests}\label{sec:chi}
The $\chi^2$ test, proposed by Karl Pearson in 1900 \citep{pearson1900x}, is among the most useful goodness-of-fit tests. For the univariate case, the range of the $n$ observations is divided into $k$ mutually exclusive classes; $O_i=n_i$ is the observed frequency in class $i$, and $p_i$ is the probability that an observation will fall into class $i$ under the null hypothesis, so that $E_i=np_i$ is the expected frequency in class $i$. The  $\chi^2$ statistic is then given by
\begin{equation}	
	\chi^2=\sum_{i=1}^{k}\frac{(n_i-np_i)^{2}}{np_i}=\sum_{i=1}^{k}\frac{(O_i-E_i)^2}{E_i},
	\label{eq:chisq}
\end{equation}
which is asymptotically $\chi_{k-1}^{2}$ under any null distribution. One disadvantage of the $\chi^2$ test is that the testing results can be substantially affected by the number and size of the $k$ classes chosen (see Section 5.2 in \cite{thode2002testing} for more details). The $\chi^2$ test is, however, not recommended as a test for univariate normality \citep{moore1986tests}, mostly because of its lack of power relative to other tests for normality. However, the test is easily adaptable to any null distribution, including those that are multivariate in nature, so that it can be used for testing MVN rather than other tests that are much more difficult to implement. As in the univariate case, the sample space is required to be partitioned into mutually exclusive classes; hence, the same problem must still be addressed, i.e., the class size and number of classes. In addition, the problem of choosing class intervals becomes much more difficult as the dimension of the sample space increases, and even in the multivariate normal case, calculating expected frequencies can be extremely difficult. Early attempts to develop extensions of $\chi^2$ test for MVN include \cite{kowalski1970performance}, \cite{moore1981chi} and \cite{mason1985re}, and a few recent studies, presented below, also focused on the chi-squared type tests for MVN.

\cite{cardoso2010multivariate} proposed a multivariate $\chi^2$ test for MVN based on the fact that the statistics
\begin{equation}
	B_i=\frac{n}{(n-1)^2}(\bX_i-\overbar{\bX})^{\top}\widetilde{\bS}^{-1}(\bX_i-\overbar{\bX}),\;i=1,\ldots,n,
	\label{eq:bi}
\end{equation}
where $\widetilde{\bS}$ is the unbiased sample covariance matrix, are each distributed exactly as $\mbox{Beta}(p/2,(n-p-1)/2)$ under $H_0$ \citep{gnanadesikan1972robust}. The authors defined $k$ equal-sized classes based on the empirical rule
\begin{empheq}[left={k\approx\empheqlbrace}]{alignat*=2}
	&\sqrt{n}, & \qquad\mbox{ if } n\leq 100, \\[-10pt]
	&5\log_{10}(n), & \qquad\mbox{ if } n>100.
\end{empheq}
The class intervals in the sample space of $B_1,\ldots,B_n$ correspond to regions partitioned from the original $p$-dimensional sample space of $\bX_1,\ldots,\bX_n$. Now, let $q_i$ be the upper $(k-i)/k\times 100\%$ quantile of the $\mbox{Beta}(p/2,(n-p-1)/2)$ distribution, then the $i$th class is defined by $\{q|q_{i-1}<q\leq q_i\}$ for $i=1,\ldots,k$, where $q_0=0$ and $q_{k}=1$. The observed frequency $O_i$ of the $i$th class is the number of values for $B_1,\ldots,B_n$ that fall within the class limit $(q_{i-1},q_i]$, and the expected frequency is simply $E_i = n/k$, $i=1,\ldots,k$. The test statistic is then calculated using Equation~(\ref{eq:chisq}), which is asymptotically distributed as $\chi_{k-1}^2$ under $H_0$.

Noticing that the above testing procedure was a $k$-dimensional multinomial goodness-of-fit test, and Pearson's $\chi^2$ statistic was used to measure the discrepancy between the observed and expected proportions, \cite{batsidis2013necessary} proposed a broader class of tests based on the power divergence family of statistics \citep{cressie1984multinomial, read2012goodness}:
\[
Z_{(\lambda)}=\begin{cases}
	\frac{2}{\lambda(\lambda+1)}\sum_{i=1}^{k}O_{i}\left\{\left( \frac{O_i}{E_i} \right)^{\lambda}-1\right\}, & \mbox{ when } \lambda\in\bbR,\lambda\neq -1, 0, \\
	2\sum_{i=1}^{k}E_i\log\frac{E_i}{O_i}, & \mbox{ when } \lambda=-1,\\
	2\sum_{i=1}^{k}O_i\log\frac{O_i}{E_i}, & \mbox{ when } \lambda=0,
\end{cases}
\]
which includes as a specific case the Pearson's $\chi^2$ statistic, Equation~(\ref{eq:chisq}), when $\lambda=1$. Here $Z_{(\lambda)}$ is also aymptotically $\chi_{k-1}^2$ under $H_0$, where $O_i$ and $E_i$ are calculated in the same way as in \cite{cardoso2010multivariate}.

Apart from formal testing procedures for MVN with explicitly defined test statistics, subjective graphical methods based on quantiles have also been proposed, such as \cite{small1978plotting}, who assessed MVN based on the plot of the points $(B_{(i)},D_{i}), i=1,\ldots,n$ with the line $y=x$, where $B_{(i)}$'s are the ordered statistics of $B_i$'s defined in Equation~(\ref{eq:bi}), and $D_i$'s are Beta order statistics using Blom's general plotting position \citep{blom1958statistical}: $\frac{i-\alpha}{n-\alpha-\beta+1},\;\;i=1,\ldots,n,$
with $\alpha=(p-2)/(2p)$ and $\beta=0.5-(n-p-1)^{-1}$. Another graphical method was proposed by \cite{srivastava1984measure}. \cite{hanusz2012new} formalized both graphical methods using explicit test statistics. For example, they formalized the testing procedure of \cite{small1978plotting} by constructing a geometric test statistic, $\textrm{SmG}$, that measures the departure of empirical points from the line $y = x$, i.e., the sum of the areas between the points $(B_{(i)},D_{i}), i=1,\ldots,n$ and the line $y=x$, as shown in their Figure~1. Large values of the statistic lead to rejection of MVN of the data. \cite{madukaife2019new} pointed out that some areas in the above test statistic may be irregular in shape, and thus may not be easily computed without the use of special computer programs. They therefore proposed a more tractable statistic based on the distances between an ordered set of the transformed observations
\[
Z_i=(\bX_i-\overbar{\bX})^{\top}\widetilde{\bS}^{-1}(\bX_i-\overbar{\bX}),\;i=1,\ldots,n,
\]
which are asymptotically distributed as $\chi_{p}^{2}$ under $H_0$, and the set of the population quantiles of the $\chi_{p}^2$ distribution. Specifically, the test statistic is
\[
G=\sum_{i=1}^{n}(Z_{(i)}-C_i)^2,
\]
where $Z_{(i)}$'s are the ordered statistics of $Z_{i}$'s, and $C_i$'s are the corresponding approximate expected order statistics, i.e., the quantiles of the $\chi_{p}^2$ distribution. Again, large values of $G$ will lead to rejection of MVN of the data. 

\cite{voinov2016new} found that the $\chi^2$ test statistic for MVN, i.e., the Nikulin-Rao-Robson (NRR) statistic, proposed in \cite{moore1981chi} is asymptotically chi-square distributed under $H_0$ if and only if the covariance matrix $\bSigma$ is a diagonal matrix. They derived the forms of the NRR statistic, $Y_{n}^{2}$, as well as its decomposition, $Y_{n}^2=U_{n}^2+S_{n}^2$, for any diagonal covariance matrix of any dimensionality $p$ (see their equations (6), (9) and (10)) and suggested a procedure for testing MVN: 1) produce the Karhunen-Lo{\`e}ve transformation of the sample data, which will diagonalize the sample covariance matrix, and 2) compute the statistics $Y_{n}^2$, $U_{n}^2$ and $S_{n}^2$ according to their equations (6), (9) and (10), respectively, based on the transformed data. Since $U_{n}^2$ and $S_{n}^2$ are asymptotically independent under $H_0$, they can be used as test statistics independently from each other.

\subsection{BHEP-type tests}
\label{sec:BHEP}
The BHEP (Baringhaus-Henze-Epps-Pulley) tests, coined by \cite{csorgHo1989consistency}, is a class of affine invariant and universally consistent tests for MVN based on the empirical characteristic function (CF). \cite{epps1983test} provided a test for univariate normality based on the empirical CF, and \cite{baringhaus1988consistent} generalized their idea to the multivariate case. \cite{henze1990class} studied the test in a more general setting to gain more flexibility with respect to the power of the test against specific alternatives. The BHEP statistic is given by
\begin{equation}
T_{n,\beta}=n\int_{\bbR^{p}}\left| \Psi_{n}(\bt)-\Psi(\bt)\right|^2 \psi_{\beta}(\bt)\mbox{d}\bt,
	\label{eq:BHEP}
\end{equation}
where $\beta>0$ is the smoothing parameter, $\Psi_{n}(\bt)=\frac{1}{n}\sum_{j=1}^{n}\exp (i\bt^{\top}\bY_{j})$ is the empirical CF of the scaled residuals $\bY_{j}, j=1,\ldots,n$, $\Psi(\bt)=\exp \left(-\|\bt\|^2/2\right)$ is the CF of $\mathcal{N}_{p}(\textbf{\textit{0}},\bI_{p})$, and the weighting function $\psi_{\beta}(\bt)=(2\pi\beta^2)^{-p/2}\exp \left(-\frac{\|\bt\|^2}{2\beta^2} \right)$ is the density of $\mathcal{N}_{p}(\textbf{\textit{0}},\beta^2 \bI_{p})$. Theoretical properties of the statistic $T_{n,\beta}$ and alternative test statistics based on the empirical CF using other functional distances have been studied by \cite{baringhaus1988consistent}, \cite{csorgHo1989consistency},  \cite{henze1990approximation}, \cite{henze1997extreme}, \cite{henze1997new} and \cite{epps1999limiting} (see Section 6 in \cite{henze2002invariant} and the references therein). Continuous interest has been shown in developing BHEP-type tests since the review paper of \cite{henze2002invariant}, as discussed below.

\cite{pudelko2005new} proposed a test statistic based on the weighted supremum distance:
\[
T_{n,r}=\sqrt{n}\sup_{\|\bt\|<r}|W_{n}(\bt)|,
\]
where $r>0$ and
\begin{equation*}
	W_{n}(\bt)=\begin{cases}
		\frac{\Psi_{n}(\bt)-\Psi(\bt)}{\|\bt\|}, & \bt\neq\textbf{\textit{0}},\\
		0,&\bt=\textbf{\textit{0}},
	\end{cases}
\end{equation*}
with $\Psi_{n}(\bt)$ and $\Psi(\bt)$ defined as above.
The asymptotic null distribution is derived as the distribution of the supremum norm of a non-stationary complex-valued $d$-dimensional Gaussian random process.

\cite{arcones2007two} proposed two BHEP-type tests based on the L{\'e}vy characterization of the normal distribution \citep{levy1977prob} and its variant. The test statistics, however, are rather complicated to compute. For example, the first test statistic is given by
\[
\widehat{D}_{n,m}=\int_{\bbR^{p}}\left| \widehat{\psi}_{n,m}(\bt)-\Psi(\bt)\right|^2 \psi_{\beta}(\bt)\mbox{d}\bt,
\]
where
\[
\widehat{\psi}_{n,m}(\bt):=\frac{(n-m)!}{n!}\sum_{(j_1,\ldots,j_m)\in I_{m}^n}\exp\left[ im^{-1/2}\bt^{\top}\left\{ \sum_{k=1}^{m}\widehat{\bSigma}_{n}^{-1/2}(\bX_{j_k}-\widehat{\bmu}_n) \right\}\right],
\]
$\widehat{\bmu}_n$ and $\widehat{\bSigma}_{n}$ are estimators of $\bmu_{F_X}$ and $\bSigma_{F_X}$, respectively, and $I_{m}^{n}=\{ (j_{1},\ldots,j_{m})\in\mathbb{N}^{m}:1\leq j_{k}\leq n,j_{k}\neq j_{l} \mbox{ if } k\neq l \}$. If $m=1$, $\widehat{\bmu}_n=\overbar{\bX}$ and $\widehat{\bSigma}_{n}=\bS$, then $\widehat{D}_{n,m}$ agrees with $T_{n,\beta}$ in Equation~(\ref{eq:BHEP}). 

\cite{henze2019new} constructed a ``moment generating function (MGF) analogue'' to the BHEP statistic $T_{n,\beta}$. The test statistic is given by
\[
\widetilde{T}_{n,\beta}=n\int_{\bbR^{p}}\{M_{n}(\bt)-m(\bt)\}^{2} \omega_{\beta}(\bt)\mbox{d}\bt,
\]
where $M_{n}(\bt)=\frac{1}{n}\sum_{j=1}^{n}\exp(\bt^{\top}\bY_{j})$ is the empirical MGF of the scaled residuals $\bY_{j}, j=1,\ldots,n$, $m(\bt)=\exp(\|\bt\|^2/2)$ is the MGF of $\mathcal{N}_p(\textbf{\textit{0}},\bI_{p})$, and $\omega_{\beta}(\bt)=\exp (-\beta\|\bt\|^2)$ with $\beta>1$ is the weighting function, which leads to a representation of $\widetilde{T}_{n,\beta}$ (see their Equation (1.4)) that is amendable to computational purposes. The authors showed that after a suitable scaling, $\widetilde{T}_{n,\beta}$ approaches a
linear combination of sample measures of multivariate skewness in the sense of \cite{mardia1970measures} and \cite{mori1994multivariate}, as $\beta\rightarrow \infty$ (see their Theorem 2.1). They also showed that $\widetilde{T}_{n,\beta}$ has a non-degenerate asymptotic null distribution only when $\beta>2$. 

\cite{henze2019characterizations} constructed a class of tests based on both the CF and the MGF. The authors generalized a characterization of univariate normal distributions in \cite{volkmer2014characterization} to the multivariate case (see their Proposition 2.1), and showed that $\bX\in\bbR^{p}$ is zero-mean normal distributed if and only if $R_{\bX}(\bt)M_{\bX}(\bt)-1=0$, where $R_{\bX}(\bt)=\mbox{Re}\{\phi_{\bX}(\bt)\}$ is the real part of the CF, $\phi_{\bX}(\bt)$, and $M_{\bX}(\bt)$ is the MGF of $\bX$. Let $R_{n}(\bt)=\frac{1}{n}\sum_{j=1}^{n}\cos (\bt^{\top}\bY_{j})$ be the empirical cosine transform, $M_{n}(\bt)=\frac{1}{n}\sum_{j=1}^{n}\exp(\bt^{\top}\bY_{j})$ be the empirical MGF of the scaled residuals $\bY_{j}, j=1,\ldots,n$, and $U_{n}(\bt)=\sqrt{n}\{R_{n}(\bt)M_{n}(\bt)-1\}$. The test statistic is given by
\[
T_{n,\gamma}=\int_{\bbR^{p}}U_{n}^{2}(\bt)\omega_{\gamma}(\bt)\mbox{d}\bt=n\int_{\bbR^{p}}\{R_{n}(\bt)M_{n}(\bt)-1\}^2 \omega_{\gamma}(\bt)\mbox{d}\bt,
\]
where $\omega_{\gamma}(\bt)=\exp(-\gamma\|\bt\|^2)$ with $\gamma>0$ is the weighting function, which leads to a computationally feasible form of $T_{n,\gamma}$ (see their Equation (3.7)). They found a simpler form if the test statistic if defined by $\widetilde{T}_{n,\gamma}=\int_{\bbR^{p}}U_{n}(\bt)\omega_{\gamma}(\bt)\mbox{d}\bt$:
\[
\widetilde{T}_{n,\gamma}=\left(\frac{\pi}{\gamma}\right)^{p/2}\sqrt{n}\left\{ \frac{1}{n^2}\sum_{j,k=1}^{n}\exp\left(\frac{\|\bY_{j}\|^2-\|\bY_k\|^2}{4\gamma}\right)\cos\left(\frac{\bY_j^{\top}\bY_{k}}{2\gamma}\right)-1 \right\}.
\]
The asymptotic null distribution of $\widetilde{T}_{n,\gamma}$ is $\mathcal{N}(0,\sigma^2)$, where $\sigma^2=2\pi^{p}(\gamma^2-0.25)^{-p/2}+2\pi^{p}(\gamma^2+0.25)^{-p/2}-4\pi^{p}\gamma^{-p}$.

\begin{table}[h!]
	\centering
	\caption{Properties of the recent tests and classical tests for MVN for i.i.d. data.} 
	\label{table1}
	\begingroup
	\setlength{\tabcolsep}{6pt} 
	\renewcommand{\arraystretch}{1.2} 
	\resizebox{\textwidth}{!}{%
		\begin{tabular}{c | ccccc}
			\hline\hline
			Test & Affine invariance & Universal consistency & Known null distribution  & Reference \\
			\hline 
			\multicolumn{2}{l}{1. Skewness and kurtosis approaches}\\
			\hline
			\textrm{MS}, \textrm{MK} & $\checkmark$ & $\text{\sffamily x}$ & $\checkmark$ &  \cite{mardia1974applications}\\
			$U, W$ & $\checkmark$ & $\text{\sffamily x}$ & $\checkmark$ & \cite{kankainen2007tests}\\
			$Z_{2,p}^{HL}$ & $\checkmark$ & $\text{\sffamily x}$ & $\text{\sffamily x}$ & \cite{thulin2014tests}\\
			$b_{2,p,q}$ & $\checkmark$ & $\text{\sffamily x}$ & $\checkmark$ & \cite{yamada2015kurtosis}\\
			$\mbox{JB}_{\mbox{\tiny BS}}$  & $\checkmark$ & $\checkmark$ & $\checkmark$ & \cite{bowman1975omnibus}\\
			$\mbox{JB}_{\mbox{\tiny DH}}$  & $\checkmark$ & $\checkmark$ & $\checkmark$ & \cite{doornik2008omnibus}\\
			$\widetilde{\textrm{LM}}$ & $\checkmark$ & $\checkmark$ & $\checkmark$ & \cite{jonsson2011robust}\\
			$\textrm{JB}_M$, $\textrm{RJB}_M$, $\textrm{RT}_M$, $\textrm{JBT}_M$ & $\checkmark$ & $\checkmark$ & $\checkmark$ & \cite{kim2016robustified}\\
			$M_1, M_2$ & $\checkmark$ & $\checkmark$ & $\checkmark$ & \cite{koizumi2009jarque}\\
			$M_3$ & $\checkmark$ & $\checkmark$ & $\checkmark$ & \cite{enomoto2012multivariate}\\
			$\textrm{MJB}_2$, $mMJB$ & $\checkmark$ & $\checkmark$ & $\checkmark$ & \cite{koizumi2014modified}\\
			\hline
			\multicolumn{2}{l}{2. Chi-squared type tests}\\
			\hline
			$\textrm{NRR}$ & $\checkmark$ & $\text{\sffamily x}$ & $\checkmark$ & \cite{moore1981chi}\\
			$\chi^2$ & $\checkmark$ & $\text{\sffamily x}$ & $\checkmark$ & \cite{cardoso2010multivariate}\\
			$Z_{(\lambda)}$ & $\checkmark$ & $\text{\sffamily x}$ & $\checkmark$ & \cite{batsidis2013necessary}\\
			$\textrm{SmG}$ & $\checkmark$ & $\text{\sffamily x}$ & $\text{\sffamily x}$ & \cite{hanusz2012new}\\
			$G$ & $\checkmark$ & $\checkmark$ & $\text{\sffamily x}$ & \cite{madukaife2019new}\\
			$Y_{n}^2, U_{n}^2, S_{n}^2$ & $\checkmark$ & $\checkmark$ & $\checkmark$ & \cite{voinov2016new}\\
			\hline
			\multicolumn{2}{l}{3. BHEP-type tests}\\
			\hline
			$T_{n,\beta}$ & $\checkmark$ & $\checkmark$ & $\checkmark$ & \cite{henze1990class}\\
			$T_{n,r}$ & $\checkmark$ & $\checkmark$ & $\checkmark$ & \cite{pudelko2005new}\\
			$\widehat{D}_{n,m}$ & $\checkmark$ & $\checkmark$ & $\checkmark$ & \cite{arcones2007two}\\
			$\tilde{T}_{n,\beta}$ & $\checkmark$ & $\checkmark$ & $\checkmark$ &  \cite{henze2019new}\\
			$T_{n,\gamma}$, $\widetilde{T}_{n,\gamma}$ & $\checkmark$ & $\checkmark$ & $\checkmark$ &  \cite{henze2019characterizations}\\
			\hline
			\multicolumn{2}{l}{4. Other generalizations of univariate normality test}\\
			\hline
			$\delta_{n,p}$ & $\checkmark$ & $\checkmark$ & $\text{\sffamily x}$ & \cite{szekely2005new}\\
			$W^{*}$ & $\checkmark$ & $\checkmark$ & $\text{\sffamily x}$ & \cite{villasenor2009generalization}\\
			$\textrm{MPI}$ & $\text{\sffamily x}$ & $\text{\sffamily x}$ & $\text{\sffamily x}$ &\cite{majerski2010approximations}\\
			$Z_{A}^{*}, Z_{A}^{**}$ & $\checkmark$ & $\checkmark$ & $\text{\sffamily x}$ & \cite{kim2018likelihood}\\
			\hline
			\multicolumn{2}{l}{5. Multiple test procedures}\\
			\hline
			$T_{n}(u)$ & $\checkmark$ & $\checkmark$ & $\text{\sffamily x}$ & \cite{tenreiro2011affine, tenreiro2017new}\\
			$T_{n,c}$ & $\checkmark$ & $\checkmark$ & $\text{\sffamily x}$ & \cite{zhou2014powerful}\\
			\hline \hline
	\end{tabular} }
	\endgroup
\end{table}
\clearpage

\newpage
\section{Simulation Study}\label{sec:simu}
In this section, we investigate the influence of spatial dependence on the measures of skewness and kurtosis for multivariate Gaussian random fields through Monte Carlo simulation studies. The results reveal that the sample skewness and kurtosis deviate from their theoretical values in the i.i.d. case as the degree of spatial dependence increases. Due to these deviations, the usual test statistics based on sample skewness and kurtosis can have a quite different asymptotic behavior under spatial dependence, so that the usual test of normality, which depends on the asymptotic property derived under the i.i.d. assumption, may be misleading for spatially correlated data. Therefore, there is a need to construct a new MVN test under spatial dependence, which is the focus of the next section.

For a multivariate random field, the cross-covariances measure the spatial dependences within individual variables as well as between distinct variables. 
For a $p$-variate random field $\bZ(\bs)=(Z_{1}(\bs),Z_{2}(\bs),\ldots,Z_{p}(\bs))^{\top},\:\bs\in\bbR^{d}$, the matrix-valued cross-covariance function of $\bZ(\bs)$ at two locations, $\bs_1\in\bbR^d$ and $\bs_2\in\bbR^d$, is defined as $\bC(\bs_1,\bs_2)=\{C_{ij}(\bs_1,\bs_2)\}_{i,j=1}^{p}$, where $C_{ij}(\bs_1,\bs_2)=\mbox{cov}\{ Z_{i}(\bs_1), Z_{j}(\bs_2)\},\: i,j=1,\ldots,p$. The covariance matrix $\bSigma=\{\bC(\bs_i,\bs_j)\}_{i,j=1}^{n}$ should satisfy the nonnegative definite condition: $\ba^{\top}\bSigma\ba\geq 0$ for any vector $\ba\in\bbR^{np}$, any spatial locations $\bs_1,\ldots,\bs_n$, and any integer $n$.
Various valid cross-covariance models have been built (see \cite{genton2015cross} for a review), and the multivariate Mat{\'e}rn model \citep{gneiting2010matern} has received a great deal of attention.

In particular, the parsimonious Mat{\'e}rn model for a stationary bivariate random field, where the cross-covariances depend on the spatial lags only, is given by
\begin{align}
	&C_{11}(\bh)=\sigma_{1}^{2}\mathcal{M}(\bh|\nu_1,\beta), \quad C_{22}(\bh)=\sigma_{2}^{2}\mathcal{M}(\bh|\nu_2,\beta),
	\label{eq:bimatern1}\\
	&C_{12}(\bh)=C_{21}(\bh)=\rho_{12}\sigma_{1}\sigma_{2}\mathcal{M}\Big(\bh|\frac{1}{2}(\nu_1+\nu_2),\beta\Big), 
	\label{eq:bimatern2}
\end{align}  
where $\sigma_{1}^2$ and $\sigma_{2}^2$ are the marginal variances, $\mathcal{M}(\bh|\nu,\beta)=\frac{2^{1-\nu}}{\Gamma(\nu)}\left(\|\bh\|/\beta\right)^{\nu}\mathcal{K}_{\nu}\left(\|\bh\|/\beta\right)$, $\nu>0$ is the smoothness parameter, $\beta>0$ is the spatial range parameter, and $\mathcal{K}_{\nu}$ is a modified Bessel function of the second kind of order $\nu$. The colocated correlation coefficient $\rho_{12}$ should satisfy the following condition for the model to be valid:
\begin{equation}
	|\rho_{12}|\leq \frac{\Gamma(\nu_{1}+\frac{d}{2})^{1/2}}{\Gamma(\nu_1)^{1/2}}\frac{\Gamma(\nu_{2}+\frac{d}{2})^{1/2}}{\Gamma(\nu_2)^{1/2}}\frac{\Gamma\{\frac{1}{2}(\nu_1+\nu_2)\}}{\Gamma\{\frac{1}{2}(\nu_1+\nu_2)+\frac{d}{2}\}}.
	\label{eq:rho}
\end{equation}

In this section, we simulate bivariate random fields defined on $[0,1]\times[0,1]\subset\bbR^{2}$ with certain cross-covariance structures, and examine the behaviors of sample skewness and kurtosis as a function of the degree of spatial dependence specified in the cross-covariance function. Specifically, we use the bivariate Mat{\'e}rn model (\ref{eq:bimatern1}) and (\ref{eq:bimatern2}) with smoothness parameters $\nu_1=\nu_2=0.5$ (Exponential) or $\nu_1=\nu_2=1$ (Whittle), and the colocated correlation coefficient $\rho_{12}$ can be either positive (e.g., $0.5$) or negative (e.g., $-0.5$) as long as it satisfies the inequality (\ref{eq:rho}). Both marginal variances are set to $1$ for simplicity. Further, the spatial dependence can be characterized by the effective range $h^{*}$, which is defined as the distance beyond which the correlation between observations is less than or equal to $0.05$ \citep{IGH:07}. We simulate the random fields at a $15\times15$ regular grid of locations over the unit square, set the effective range $h^{*}\in\{0.1,0.12,0.14,\ldots,0.88,0.9\}$, implying an increasing degree of spatial dependence, and solve the following equations: 
\begin{equation*}
	R(h^{*})=\exp\left(-\frac{h^{*}}{\beta}\right)=0.05 \:\: \mbox{ (Exponential) \quad or } \quad R(h^{*})=\frac{h^{*}}{\beta}\mathcal{K}_{1}\left(\frac{h^{*}}{\beta}\right)=0.05  \:\:\mbox{ (Whittle) }
\end{equation*}
to get the values of the spatial range parameter $\beta$. We simulate 200 replicates for each combination of parameters. In order to see the pure effect of spatial dependence determined by $h^{*}$ or the induced parameter $\beta$, in each simulation we simulate a standard multi-normal random vector $\mathbf{e}$ and fix it, and then impose the covariance matrix on it. Specifically, to simulate a bivariate random field $\bZ(\bs)=(Z_{1}(\bs),Z_{2}(\bs))^{\top}$ at a regular grid of $n$ locations, we first stack the variables in a long vector $\bZ=(\bZ_{1}^{\top},\bZ_{2}^{\top})^{\top}=(Z_{1}(\bs_1),\ldots,Z_{1}(\bs_n),Z_{2}(\bs_1),\ldots,Z_{2}(\bs_n))^{\top}$, then simulate and fix a standard multi-normal random vector $\boldsymbol{e}\in\bbR^{2n}$, and get the values of $\bZ$ by $\bZ=\bSigma^{1/2}(\boldsymbol{\theta}(h^{*}))\boldsymbol{e}\in\bbR^{2n}$,
for each combination of parameters $\boldsymbol{\theta}$ that depends on the effective range $h^{*}$, where $\bSigma^{1/2}$ is the square root of $\bSigma=\begin{bmatrix}
	\bSigma_{11} & \bSigma_{12} \\
	\bSigma_{21} & \bSigma_{22} 
\end{bmatrix}$, the covariance matrix of $\bZ$, with $\bSigma_{11}$ and $\bSigma_{22}$ being the auto-covariance matrices for $\bZ_{1}$ and $\bZ_{2}$, respectively, and $\bSigma_{12}=\bSigma_{21}^{\top}$ being the cross-covariance matrix between $\bZ_1$ and $\bZ_2$. By doing this, we can eliminate the effect of randomness coming from $\mathbf{e}$ and isolate the effect of changing the parameters, particularly the degree of spatial dependence, in the covariance function.

Following these procedures, we thus have 200 sample skewness and kurtosis for each level of spatial dependence (i.e., the effective range $h^{*}$ or the correlation parameter $\rho_{12}$) that is specified in the covariance structure. We then summarize the 200 curves of sample skewness and kurtosis as a function of $h^{*}$ or $\rho_{12}$ by functional boxplot \citep{sun2011functional}, which is an extension of the classical boxplot for visualizing functional data. A functional boxplot displays three descriptive statistics: the median curve, the envelope of the 50\% central region, and the maximum non-outlying envelope \citep{sun2011functional}. Outliers are detected as exceeding 1.5 times the 50\% central region, similarly to classical boxplots. 

\begin{figure}
	\centering
	\includegraphics[width=0.9\textwidth]{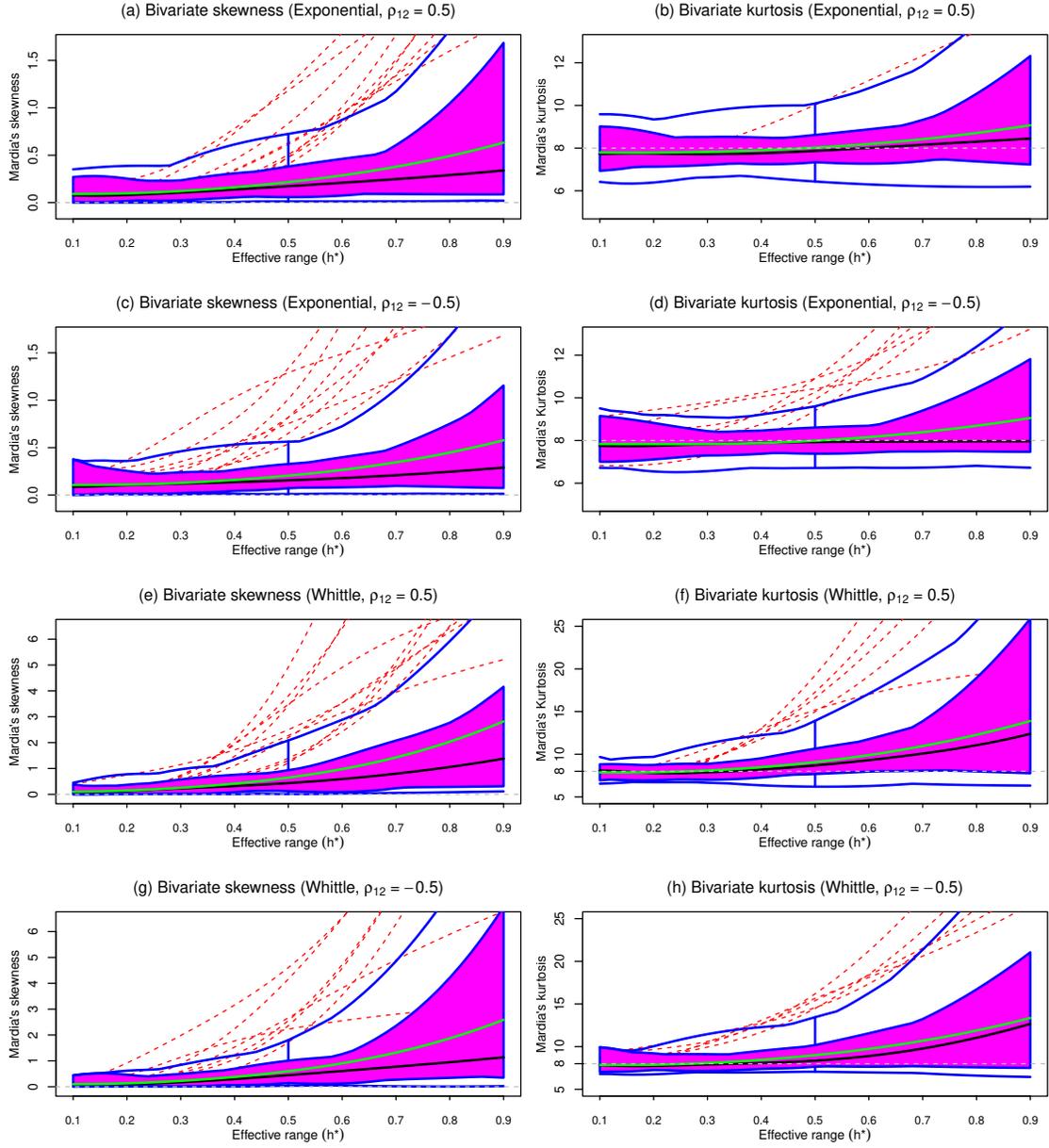}
	\caption{\baselineskip=16pt Functional boxplot of the Mardia's sample skewness and kurtosis of the bivariate Gaussian random field in $[0,1]\times[0,1]$ as a function of the effective range $h^{*}$ for (a)-(d) the Exponential and (e)-(h) the Whittle covariance functions. The green curve is the point-wise mean curve, the black curve is the median curve, the purple shaded region is the envelope of the 50\% central region,  the outer blue curves represent the maximum non-outlying envelope, and the red dashed curves are detected outliers. The theoretical values of Mardia's measures of skewness (i.e., $\beta_{1,2}=0$) and kurtosis (i.e., $\beta_{2,2}=8$) for a bivariate normal distribution are indicated by gray dashed lines.}
	\label{fig:biv2d}
\end{figure}

Figure~\ref{fig:biv2d} 
shows the functional boxplots of the Mardia's sample skewness and kurtosis of the bivariate Gaussian random field on $\bbR^2$ as a function of the effective range $h^{*}$. Recall that Mardia's measure of multivariate skewness is always positive. We find that the sample skewness and kurtosis increase as the effective range increases, and the smoother the field, the larger the influence from spatial dependence. The difference between the cases where $\rho_{12}>0$ and where $\rho_{12}<0$ is small if we compare, e.g., (a) with (c) or (b) with (d).

\section{The New Test for MVN Under Spatial Dependence}\label{sec:newtest}
\subsection{Construction of the new test}
The results from the simulation study in the previous section suggest that the dependence in spatial data should be appropriately accounted for in the tests for MVN based on sample skewness and kurtosis measures. Otherwise, the un-adjusted tests may lead to conservative decisions on assessing the Gaussianity in the data; that is, data from a Gaussian random field with spatial dependence tend to be detected as being non-Gaussian. 
\cite{horvath2020testing} proposed a JB-type test to address this problem for the univariate case. Assume that the spatial dataset $\{X({\bs}_1),X({\bs}_2),\ldots,X({\bs}_n)\}$, where $\{{\bs}_1,{\bs}_2,\ldots,{\bs}_n\}\in\mathbb{Z}^{d}$ are locations in the $d$-dimensional space with integer coordinates, is from a strictly stationary Gaussian spatial moving average process under the $H_0$:
\begin{equation}
X(\bs)=\mu+\sum_{\bt\in\mathbb{Z}^{d}}a(\bt)\varepsilon(\bs-\bt), \;\;\bs\in\mathbb{Z}^{d},
\label{eq:sma}
\end{equation}
where the innovations $\varepsilon(\bs), \bs\in\mathbb{Z}^{d}$ are i.i.d. from $\mathcal{N}(0,1)$, and the constants $a(\bs), \bs\in\mathbb{Z}^{d}$, satisfy $\sum_{\bs\in\mathbb{Z}^{d}}|a(\bs)|^2<\infty$. 
The JB-type test statistic is 
\begin{equation}
\textrm{JB}^* = \frac{S_n^2}{\hat{\phi}_\textrm{S}^2} + \frac{K_n^2}{\hat{\phi}_\textrm{K}^2},
\label{eq:JB1}
\end{equation}
where $S_n$ and $K_n$ are sample skewness and kurtosis of the standardized observations, respectively, and $\hat{\phi}_\textrm{S}^2$ and $\hat{\phi}_\textrm{K}^2$ are consistent estimators of the asymptotic variances of $S_n$ and $K_n$, respectively. \cite{horvath2020testing} defined the kernel estimators, $\hat{\phi}_\textrm{S}^2$ and $\hat{\phi}_\textrm{K}^2$, as 
\begin{align}
\hat{\phi}_\textrm{S}^2&=6\sum_{\bh}\omega_{\boldsymbol{b}}(\bh)\widehat{C}^{3}(\bh):=6\sum_{l=1}^{d}\sum_{|h_l|\leq b_l}\left\{\prod_{l=1}^{d}k\left(\frac{h_l}{b_l}\right) \right\}\widehat{C}^{3}(\bh), \label{eq:varianceS}\\
\hat{\phi}_\textrm{K}^2&=24\sum_{\bh}\omega_{\boldsymbol{b}}(\bh)\widehat{C}^{4}(\bh):=24\sum_{l=1}^{d}\sum_{|h_l|\leq b_l}\left\{\prod_{l=1}^{d}k\left(\frac{h_l}{b_l}\right) \right\}\widehat{C}^{4}(\bh),\label{eq:varianceK}
\end{align}
where $\widehat{C}(\bh)$ is the sample auto-covariance function for the standardized observations with spatial lag $\bh=(h_1,\ldots,h_d)^{\top}$; $k(\cdot)$ is a univariate kernel and $\{b_1,\ldots,b_d\}$ are smoothing bandwidths, satisfying some regularity conditions. The spatial dependence in the data is accounted for in $\widehat{C}(\bh)$, and the kernel smoothing method is used to establish consistency of the asymptotic variance estimators. Under $H_0$, the statistic JB$^*$ is asymptotically $\chi^2_2$.
 
To develop a test for the multivariate case, we adopt the union-intersection testing approach originally proposed by \cite{roy1958some}. The union-intersection principle can be formulated as follows.
Suppose we have a $p$-variate spatial dataset $\mathcal{X}=\{\bX({\bs}_1),\bX({\bs}_2),\ldots,\bX({\bs}_n)\}$, where $\{{\bs}_1,{\bs}_2,\ldots,{\bs}_n\}$ are $n$ spatial locations, $\bX(\bs_i)=(X_1(\bs_i),X_2(\bs_i),\ldots,X_p(\bs_i))^{\top}$ is the vector of $p$ variables at location ${\bs}_i, i=1,\ldots,n$. Note that the hypothesis $H_0: F_{\bX}\in \mathcal{N}_p$ holds true exactly if and only if the projection $\boldsymbol{a}^{\top}\bX$ has a univariate normal distribution for all vectors $\boldsymbol{a}\in\mathbb{R}^{p}$. For each $\boldsymbol{a}\in\mathbb{R}^{p}$, we construct a test $H_{\boldsymbol{a}}: \boldsymbol{a}^{\top}\bX$ is normal against the alternative $H_{\boldsymbol{a}}^{c}: \boldsymbol{a}^{\top}\bX$ is not normal, with acceptance region $\mathcal{A}_{\boldsymbol{a}}$ and rejection region $\mathcal{R}_{\boldsymbol{a}}$. Then the union-intersection test identifies the acceptance region for $H_0: F_{\bX}\in \mathcal{N}_p$ as $\mathcal{A}=\bigcap_{\boldsymbol{a}\in\mathbb{R}^{p}} \mathcal{A}_{\boldsymbol{a}}$, and the rejection region as $\mathcal{R}=\mathcal{A}^{c}=\bigcup_{\boldsymbol{a}\in\mathbb{R}^{p}} \mathcal{R}_{\boldsymbol{a}}$; that is, the union-intersection test does not reject $H_0$ exactly if $H_{\boldsymbol{a}}$ is not rejected for all $\boldsymbol{a}\in\mathbb{R}^{p}$, and rejects $H_0$ if $H_{\boldsymbol{a}}$ is rejected for at least one vector $\boldsymbol{a}\in\mathbb{R}^{p}$.

For a fixed $\boldsymbol{a}\in\mathbb{R}^{p}$, the projected sample $\mathcal{X}_1=\{\boldsymbol{a}^{\top}\bX({\bs}_1),\boldsymbol{a}^{\top}\bX({\bs}_2),\ldots,\boldsymbol{a}^{\top}\bX({\bs}_n)\}$ is a univariate spatial dataset, and thus we can apply the method in \cite{horvath2020testing} to test ``$H_{\boldsymbol{a}}: \boldsymbol{a}^{\top}\bX$ is normal'' based on the new sample, under the following assumption. 
\begin{Assumption}
Assume that under $H_0$, the observations $\mathcal{X}=\{\bX({\bs}_1),\bX({\bs}_2),\ldots,\bX({\bs}_n)\}$ follow
a multivariate Gaussian spatial moving average (or kernel convolution) process:
\begin{equation}
X_{l}(\bs)=\mu_l+\sigma_{l}\sum_{\bt\in\mathbb{Z}^{d}} k_{l}(\bs-\bt)\omega(\bt), \;\;\bs\in\mathbb{Z}^{d},\;l=1,\ldots,p,
\label{eq:kernel}
\end{equation}
where $\mu_l$ is the unknown mean, $\sigma_l$ is the unknown standard deviation, $k_l(\cdot),l=1,\ldots,p$, is a set of $p$ square integrable kernel functions on $\bbZ^{d}$ with $k_l(\textbf{\textit{0}}) = 1$, and $\omega(\cdot)$ is a zero-mean, unit-variance Gaussian random field on $\bbZ^{d}$ with a certain correlation function $\rho$. 
\end{Assumption}
Assumption 1 implies that under $H_0$, the linear combination $\boldsymbol{a}^{\top}\bX$, for each $\boldsymbol{a}\in\mathbb{R}^{p}$, is from a strictly stationary Gaussian spatial moving average process as defined in Equation~\eqref{eq:sma}, so that the test of \cite{horvath2020testing} can be applied. Under Assumption 1, $\bX$ is from a stationary multivariate Gaussian random field with the associated $p\times p$ matrix-valued cross-covariance function $C(\bs,\bs')$ having $(l,l')$ entry 
\begin{equation*}
(C(\bs,\bs'))_{ll'}=\sigma_{l}\sigma_{l'}\sum_{\bt\in\mathbb{Z}^{d}}\sum_{\bt'\in\mathbb{Z}^{d}} k_{l}(\bs-\bt)k_{l'}(\bs'-\bt')\rho(\bt-\bt').
\end{equation*}  
The kernel convolution technique \citep{gelfand2010multivariate} in Assumption 1 is a well-known approach for creating rich classes of stationary processes \citep{bernardo2003markov}. Therefore, our new test for MVN can be applied to spatial datasets with this big class of dependence structures.

Now, denote the JB-type test statistic for each $H_{\boldsymbol{a}}$ as $\mathrm{JB}_{\boldsymbol{a}}$, computed from Equation~\eqref{eq:JB1} based on the univariate sample $\mathcal{X}_1=\{\boldsymbol{a}^{\top}\bX({\bs}_1),\boldsymbol{a}^{\top}\bX({\bs}_2),\ldots,\boldsymbol{a}^{\top}\bX({\bs}_n)\}$. Suppose that the corresponding acceptance region is $\mathcal{A}_{\boldsymbol{a}}=\{\mathcal{X}_1:\;\mathrm{JB}_{\boldsymbol{a}}\leq c\}$ and the rejection region is $\mathcal{R}_{\boldsymbol{a}}=\{\mathcal{X}_1:\;\mathrm{JB}_{\boldsymbol{a}}> c\}$, where $c$ is a properly chosen constant (critical value) that does not depend on $\boldsymbol{a}$. Then the union-intersection test does not reject $H_0$ exactly if $\max_{\boldsymbol{a}\in\mathbb{R}^{p},\boldsymbol{a}\neq \mathbf{0}}\mathrm{JB}_{\boldsymbol{a}}\leq c$.
The critical value $c$ for the test must be determined by the distribution of the statistic $\max_{\boldsymbol{a}\in\mathbb{R}^{p},\boldsymbol{a}\neq \mathbf{0}}\mathrm{JB}_{\boldsymbol{a}}$, which is difficult to obtain in the current setting. In fact, this union-intersection test consists of infinitely many univariate tests. In practice, we can randomly select a large number of vectors, $\boldsymbol{a}_1,\ldots,\boldsymbol{a}_K\in\mathbb{R}^{p}$, and do multiple testing; if at least one test $H_{\boldsymbol{a}}$ is rejected, then $H_0$ is also rejected; otherwise, if all tests $H_{\boldsymbol{a}_1},\ldots,H_{\boldsymbol{a}_K}$ are not rejected, then this provides an evidence of not rejecting $H_0$. The number of tests, $K$, can be chosen as large as feasible for computation. In order to have a certain significance level $\alpha$ for the original test, the individual univariate tests cannot have the same level \citep{flury2013first}. Suppose that each test has a level $\alpha$, then the chance of a false rejection of the null for each test is $\alpha$, but the chance of at least one false rejection is much higher. In order to control the False Discovery Rate (FDR), which is the expected proportion of false rejections, the multiple testing procedure can be conducted based on the Benjamini-Hochberg (BH) method \citep{benjamini1995controlling}. Specifically, denote the ordered p-values for the $K$ univariate tests as $P_{(1)},\ldots,P_{(K)}$, and $R=\max\{i:P_{(i)}<\frac{i\alpha}{K}\}$. The BH rejection threshold is defined as $T=P_{(R)}$, and the hypothesis $H_{\boldsymbol{a}_i}$ is rejected if $P_i\leq T$. If this procedure is applied, then it can be shown that $\mathrm{FDR}\leq \alpha$. Since the new test is a JB-type test, it is affine invariant and universally consistent.

\subsection{Type I error and empirical power of the new test}
In this section, we assess the type I error and empirical power of the new test via Monte-Carlo simulations with various configurations of the degree of spatial dependence.  

To assess the type I error (or empirical size) of the new test, we first simulate a zero-mean $p$-variate Gaussian random field on $\bbZ^2$ (i.e., $d=2$, most commonly encountered in spatial applications) from the spatial moving average (kernel convolution) process of Equation~(\ref{eq:kernel}). Specifically, each variable is generated from the spatial moving average model defined in \cite{haining1978moving}, located on the points of a rectangular square lattice $\bbZ^2$:
\begin{equation}
X_{l}(i,j)=\theta_{l}\{e(i-1,j)+e(i+1,j)+e(i,j-1)+e(i,j+1)\}+e(i,j), \;\;l=1,\ldots,p,
\label{eq:model2}
\end{equation}
where $i$ and $j$ are integers satisfying $1\leq i \leq M$ and $1\leq j \leq N$, $e(\cdot,\cdot)$ is a zero-mean, unit-variance Gaussian process on $\bbZ^2$ with some correlation function $\rho$, and $e(i,0)=e(0,j)=e(0,0)=0$ for all $1\leq i \leq M$ and $1\leq j \leq N$. When $|\theta_l|\leq 1/4$, this model is invertible to the following first-order quadrilateral autoregressive random field:
\begin{equation*}
X_{l}(i,j)=\theta_{l}\{X(i-1,j)+X(i+1,j)+X(i,j-1)+X(i,j+1)\}+e(i,j), \;\;l=1,\ldots,p,
\end{equation*}
which has been a preoccupation for the study of finite random fields within geography as a model for spatial dependence \citep{haining1978moving}. Equation~(\ref{eq:model2}) is a special case of the spatial kernel convolution process of Equation~(\ref{eq:kernel}), where the kernels are functions taking the form of a constant height over a bounded rectangle and zero outside. To investigate the performance of the new test for different degrees of spatial dependence, we set the correlation function $\rho$ of the process $e(\cdot,\cdot)$ as the exponential correlation that has been used in Section~\ref{sec:simu}, with varying effective ranges. Without loss of generality, suppose that the $K$ vectors $\boldsymbol{a}_1,\ldots,\boldsymbol{a}_K\in\mathbb{R}^{p}$ all have norm 1, and they are chosen as $\boldsymbol{a}_i=(\cos (\theta_{1i}), \ldots, \cos (\theta_{pi}))^{\top}$, where $\theta_{1i},\ldots,\theta_{pi}$ are the $p$ coordinate direction angles in the polar coordinate system, each drawn from a uniform distribution in $[0,2\pi]$. 

Based on the above settings, we consider the bivariate case (i.e., $p=2$), set $\theta_1=1/5,\;\theta_2=-1/5$, simulate the random field at an $N\times N$ regular grid of locations over the unit square $[0,1]^2$, and vary the effective ranges, $h^*$, of the process $e(\cdot,\cdot)$ in $[0.1,0.9]$ by steps of $0.02$. For each level of the spatial dependence indicated by $h^*$, we use $1,000$ replications for the data generating and testing procedure, and the type I error is approximated by the relative frequency of null hypothesis rejection. The null hypothesis, $H_0$, is rejected when at least one of the $K$ univariate hypotheses based on the projection data is rejected using the BH method. The kernel function in the univariate test statistics for projection data is chosen as the Bartlett kernel defined as $k(t)=(1-|t|)I\{|t|\leq 1\}$ with the bandwidth $b=\lfloor{4(N/100)^{2/9}}\rfloor$; this selection of kernel and smoothing bandwidth is also used in \cite{horvath2020testing}, and it works well for our purpose. For comparison, we also apply several tests for MVN that do not account for the spatial dependence in the data, i.e., Mardia's tests, $\textrm{MS}$ and $\textrm{MK}$, defined in Equation~(\ref{eq:MS}), and the test of \cite{doornik2008omnibus}, $\mbox{JB}_{\mbox{\tiny DH}}$, defined in Equation~(\ref{eq:JB_DH}). 

To assess the empirical power of the new test, we simulate data from the non-Gaussian sinh-arcsinh (SAS) transformed multivariate Mat{\'e}rn random field defined in \cite{yan2020multivariate}. Specifically, we obtain the non-Gaussian data using the element-wise and inverse SAS transformation \citep{jones2009sinh} on the data from Gaussian random fields, i.e., the data used above for assessing the type I error. The corresponding transformation parameter setting is $(0.5,0.5)$ for the first variable and $(0.3,0.5)$ for the second variable, both have positive skewness and heavier tails than the normal distribution. Again, we use the same kernel function as above, and the empirical power is approximated by the proportion of null hypothesis rejection.

\begin{figure}[h!]
\centering
\includegraphics[width=\linewidth]{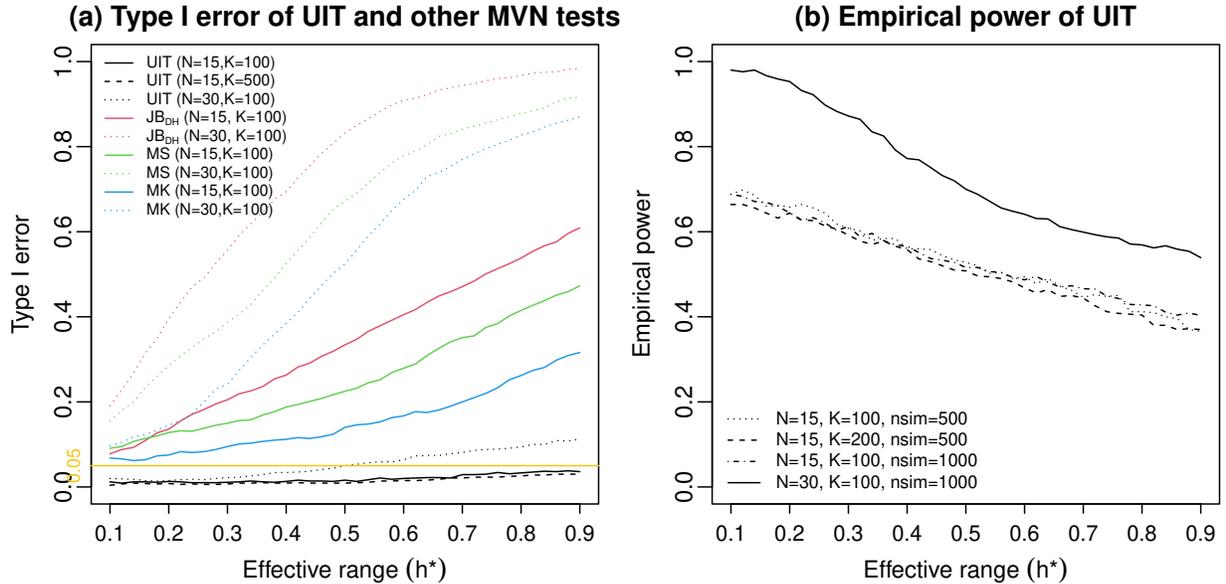}
\caption{\baselineskip=16pt (a) Type I error, as a function of the spatial dependence indicated by the effective range, of the new test (UIT, the union-intersection test) for MVN under spatial dependence (black curves) and three MVN tests for i.i.d. data (colored curves) for $N=15$ (solid curves) and $N=30$ (dotted curves), based on $1,000$ simulations for the nominal significance level of $\alpha=5\%$ (the orange horizontal line). $\mbox{JB}_{\mbox{\tiny DH}}$ (in red) represents the test of \cite{doornik2008omnibus}, and $\textrm{MS}$ (in green) and $\textrm{MK}$ (in blue) represent the tests of \cite{mardia1970measures}. The black solid curve represents the type I error of UIT for $N=15$ and $K=100$, the black dashed curve represents that for $N=15$ and $K=500$, and the black dotted curve represents that for $N=30$ and $K=100$. (b) Empirical power of the new test, UIT, as a function of the spatial dependence indicated by the effective range, for the nominal significance level of $\alpha=5\%$ for different values of $N$, $K$ and the number of simulations denoted by ``nsim''. }
\label{fig:errorpower}
\end{figure}

Figure~\ref{fig:errorpower}(a) shows the results of type I errors of our new test (UIT, union-intersection test), compared to three MVN tests for i.i.d. data for different values of $N$ and $K$ based on $1,000$ simulations. The probability of the type I error should, by any statistical test, be bounded upwards by the nominal level of significance; otherwise, the test cannot be used for the given purpose. On the other hand, a type I error far smaller than a chosen $\alpha$ is indicative of a test with low power, but does not disqualify the procedure for testing.  From Figure~\ref{fig:errorpower}(a), we can see that when $N=15$, the type I error of our new test (the black solid curve) is bounded below and not too far from the nominal significance level of $\alpha=5\%$ for all levels of spatial dependence, while the type I errors of the three MVN tests for i.i.d. data (the solid colored curves) are all severely inflated and increase as the spatial dependence gets stronger. Note that the black solid curve (with $N=15$ and $K=100$) is very close to the black dashed curve (with $N=15$ and $K=500$), indicating that $K=100$ is a large enough number of projections for the UIT test. When $N=30$, the type I error of our new test (the black dotted curve) increases as the effective range $h^*$ increases, and is slightly inflated when $h^*>0.5$, i.e., under strong spatial dependence; in contrast, all the three MVN tests for i.i.d. data exhibit inflated type I errors, even more severely than the case when $N=15$ and much higher than the type I error of the UIT test. The slightly inflated type I error of the UIT test for $N=30$ and $h^*>0.5$ is probably caused by the strong spatial dependence in the unit square, which cannot be accurately accounted for in the asymptotic variance estimators expressed by Equations~\eqref{eq:varianceS} and \eqref{eq:varianceK}. The results from Figure~\ref{fig:errorpower}(a) indicate that the MVN tests for i.i.d. data cannot be used for spatially correlated data, since they have severely inflated type I errors especially for data with strong dependence, whereas our new test can be used for spatially correlated data, and it only becomes problematic when the spatial dependence is very strong. 

Figure~\ref{fig:errorpower}(b) shows the empirical powers of our new test, UIT, for different values of $N$, $K$ and number of simulations. When $N=15$, the empirical power is not much affected by $K$ (the number of projections) and ``nsim'' (the number of simulations), since the three non-solid curves are close to each other. When $N=30$, the empirical power (shown in black solid curve) is much higher than those in the case of smaller sample size, $N=15$. In addition, all power curves go down as the effective range $h^*$ increases; moreover, when $N=30$, the power is close to one when $h^*$ is small. The results from Figure~\ref{fig:errorpower} suggest that our new test would perform best in terms of type I error and empirical power when the sample size is large and the spatial dependence is not very strong. To give a more comprehensive picture for the power performance of our new testing procedure, more investigations are needed by considering a variety of alternative non-Gaussian distributions.

\section{Wind Data Application}\label{sec:app}
\begin{figure}[h!]
	\centering
	\includegraphics[width=0.9\linewidth]{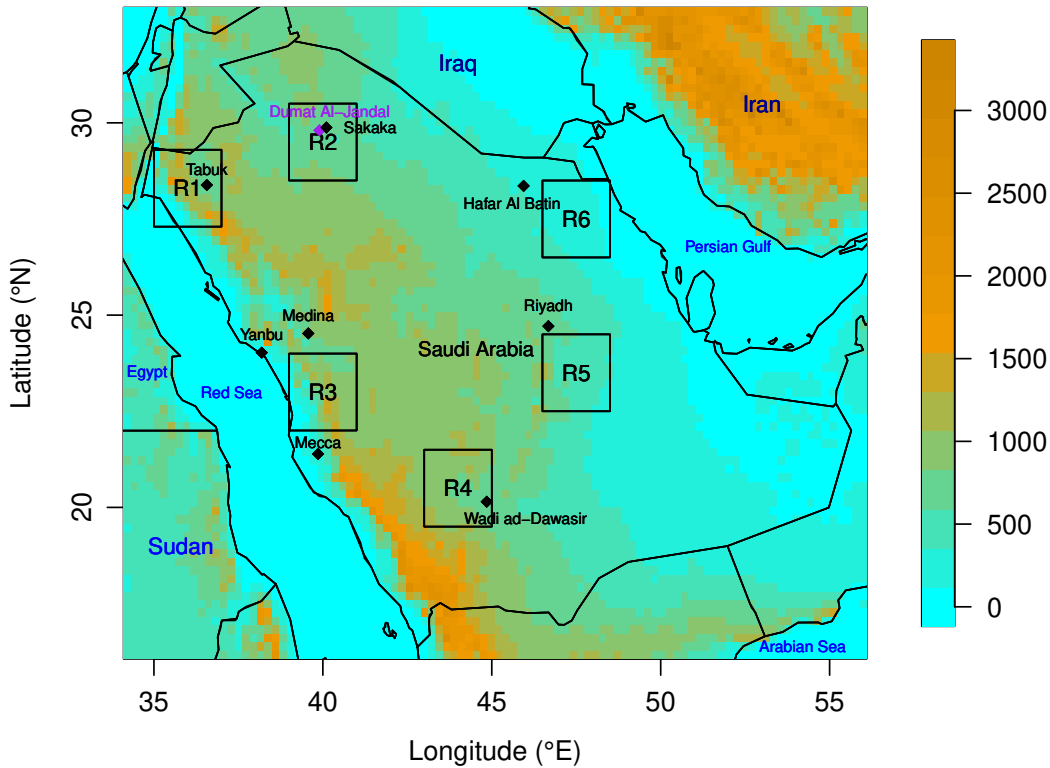}
	\vspace{-5mm}
	\caption{\baselineskip=16pt Six selected regions (denoted as R1–R6) over Saudi Arabia for testing  bivariate normality of spatial data. The color shading indicates terrain elevation (in meters).}
	\label{fig:regions}
\end{figure}

In this section, we present a data application using our new multivariate normality test for spatial data. The raw gridded data are daily U (zonal velocity) and V (meridional velocity) wind speed components during 1976-2005 over the Arabian Peninsula from the publicly available MENA CORDEX dataset \citep{ZH:17}. We use the fourth simulated historical run with a spatial resolution of $0.22\degree \times 0.22\degree$ (latitude$\times$longitude), which has been identified in \cite{Chen:18} as having the highest skill in capturing the spatio-temporal variability of reanalysis data. Following the common practice in the literature (e.g., \cite{Chen:21}), we apply the square root transform to the wind components, which stabilizes the variance over space and makes the marginal distributions approximately normal. Furthermore, in order to avoid modeling the complex seasonality in the data, we investigate the monthly average data over transformed U and V wind components in July during 30 years from 1976 to 2005. Finally, to make the spatial data approximately stationary, we deduct the long term averages from the monthly mean winds following \cite{horvath2020testing}, yielding monthly anomalies (residuals) of U and V components. Based on the pre-processed bivariate spatial data, we then test the bivariate normality over six small regions where local stationarity can be assumed, instead of the whole region where different topographies lead to spatial patterns and nonstationarity. The six regions (referred to as R1–R6 in Figure~\ref{fig:regions}) are selected similarly to those in \cite{Chen:18}. In each region, we have $12\times 12=144$ spatial locations in each of the 30 years of bivariate wind data.

We apply our new MVN test (UIT) designed for gridded stationary spatial data as well as three MVN tests designed for i.i.d. data (i.e., $\mbox{JB}_{\mbox{\tiny DH}}$, $\textrm{MS}$ and $\textrm{MK}$) at the nominal significance level of $\alpha=5\%$. Table~\ref{tab:real} shows the proportion of rejections on bivariate normality among the 30 years of July anomalies of wind U and V components over the six selected regions. We can see that in most cases, our UIT test has smaller proportions of rejections; that is, it suggests bivariate normality more often than the other three tests. In Regions 4 and 6 in particular, the UIT test does not reject normality for all 30 years, while the $\mbox{JB}_{\mbox{\tiny DH}}$ and $\textrm{MS}$ tests rejects normality for almost all 30 years. Also in Regions 1 and 2, our UIT test rejects normality in only a small number of years, while the $\mbox{JB}_{\mbox{\tiny DH}}$ and $\textrm{MS}$ tests rejects normality for all 30 years. These results imply that the MVN tests designed for i.i.d. data are usually too conservative when applied to spatially correlated data; that is,  data from a Gaussian random field with spatial dependence tend to be detected as being non-Gaussian. The MK test rejects normality less often than the UIT test in a few cases (i.e., in Regions 1, 2 and 5); this can happen since the MK test is a directed test which only considers departure from multivariate normality revealed by kurtosis, and failure to reject the null hypothesis does not necessarily imply normality, as there might be departures from normality in other ways.

\begin{table}
	\tabcolsep4.5pt
	\caption{\baselineskip=16pt Proportion of rejections of the bivariate normality hypothesis among 30 years during 1976-2005 for July anomalies of wind U and V components over the six selected regions in Saudi Arabia.}
	\vspace{-5mm}
	\begin{center}
	\begin{tabular}{c|cccccc}
	\hline\hline
	Test	& Region 1 & Region 2 & Region 3 & Region 4 & Region 5 & Region 6 \\
	\hline 
	$\mbox{UIT}$ & 0.3 & 0.4 & 0.23 & 0 & 0.53 & 0 \\
	$\mbox{JB}_{\mbox{\tiny DH}}$ & 1 & 1 & 0.83 & 1 & 1 & 0.97 \\
	$\mbox{MS}$ & 1 & 1 & 0.87 & 1 & 0.83 & 1\\
	$\mbox{MK}$ & 0.2 & 0.27 & 0.67 & 0.37 & 0.4 & 0.7\\	
	\hline\hline
	\end{tabular}
\end{center}	
\label{tab:real}
\end{table}

\section{Discussion}\label{sec:disc}
In this work, we reviewed the recent development of tests for multivariate normality for i.i.d. data, with emphasis on the skewness and kurtosis approaches. Based on simulation studies, we showed that when there exists spatial dependence in the data, the multivariate sample skewness and kurtosis measures proposed by \cite{mardia1970measures} deviate from their theoretical values under Gaussianity due to dependence, and some of the tests designed for i.i.d. data exhibit inflated type I error; the deviation and type I error increases as the spatial dependence increases. Extending the work of \cite{horvath2020testing} to the multivariate case, we then proposed a new JB-type test for multivariate normality for spatially correlated data, based on the union-intersection test approach. 
The new test has a good control of the type I error, and it is inappropriate only when the spatial dependence in the data is very strong. 
In addition, the new test has a fairly high empirical power at all levels of spatial dependence, especially for large sample sizes.

Our new test is constructed under the stationarity assumption, which should be validated before applying our test. The test for spatial stationarity proposed by, e.g., \cite{F:05} and \cite{JG:12}, can be used to check if some marginal spatial processes are nonstationary, and graphical tools such as contour plots can be used to identify possible nonstationary patterns in the cross-covariance functions. If the original data are detected as nonstationary, it is a common practice to transform them into stationarity, using the deformation approach proposed by, e.g., \cite{Schmidt:03} and \cite{Fouedjio:15}. One can also fit a nonstationary regression model (see, e.g., \cite{Schabenberger:05}) which captures most of the nonstationary features in the data, so that the residuals to be tested remain stationary.

The new test serves as a simple and useful diagnostic tool: if the null is not rejected, it lends confidence in the applications of various methodologies based on the multivariate normality assumption; if the null is rejected, it provides a caution on the validity of conclusions, and necessary pre-processing procedures may be needed before applying the methodologies, or alternative non-Gaussian methods should be considered, as illustrated in the next paragrah.

The rejection of the null hypothesis only means that the current multivariate data cannot be treated as realizations from a multivariate stationary Gaussian random field. To reveal a clearer picture of the multivariate data, the univariate normality test proposed by \cite{horvath2020testing} can be applied to each component of the variables, and the new multivariate test for spatial data we proposed can be applied to subsets of the variables, to check if some marginal processes are not normal. If that is the case, then we may need data transformations (such as the log, power or square root transformation) in order to approximately have a Gaussian process. Of course, all marginals being normal does not mean being jointly normal, so these marginal transformations may only partly help; in this case, we should be aware of the effect of conducting the current statistical procedures under the violated Gaussian assumption, and consider to switch to non-Gaussian methods (e.g., \cite{XG:17} and \cite{yan2020multivariate}).

Also note that when the sample size is large, the estimation of the auto-covariance function, i.e., $\widehat{C}(\bh)$, in Equations~\eqref{eq:varianceS} and \eqref{eq:varianceK} can be computationally prohibitive. One solution is that we can fit a parametric covariance model (such as the Mat{\'e}rn model) for $C(\bh)$, and obtain $\widehat{C}(\bh)$ by using the software \texttt{ExaGeoStat} \citep{ALSGK:18}, which allows for exact maximum likelihood estimation with dense full covariance matrices, using high performance computations. In addition, various approximation methods for large spatial datasets have also been proposed to reduce the computational burden; recent reviews include \cite{Sun:12}, \cite{Heaton:19} and \cite{Huang:21}.

One limitation of the new test, similarly to the univariate \cite{horvath2020testing} test, is that it can only be used for spatial data on a regular grid. Tests for data at irregular spatial locations need to be developed, but this can be challenging because the tests would be difficult to be justified asymptotically. Nevertheless, our proposed test can be used in various applications based on the abundant gridded data simulated from reanalysis products, General Circulation Model (GCM) experiments, Regional Climate Model (RCM) experiments or Numerical Weather Prediction (NWP) models. 

As we have mentioned in Section \ref{sec:review}, 
a way to construct multivariate JB-type tests is to combine multivariate skewness and kurtosis measures. Therefore, it would be an interesting topic to propose a JB-type test for MVN under spatial dependence that combines Mardia's multivariate skewness and kurtosis measures. Simulations in this study show that the un-adjusted tests based on Mardia's measures are misleading if applied to a spatial dataset. To account for the spatial dependence, we need to derive the asymptotic variances of the multivariate skewness and kurtosis of the scaled residuals under some kind of dependence structure, which is a non-trivial task. In addition, we need to construct consistent estimators of the asymptotic variances, and establish the asymptotic properties (limiting null distribution, etc.) of the new test. These are left for future work.


\bigskip
\begin{center}
{\large\bf SUPPLEMENTARY MATERIAL}
\end{center}

\begin{description}

\item[Title:] Review of other recent MVN tests for i.i.d. data. (PDF file)

\item[R-codes:] The R codes related to this article can be found online at the github repository: 
\texttt{https://github.com/wanruofenfang123/MVNtest\_SpatialDependence}


\end{description}


\baselineskip=20pt

\newpage
\spacingset{1.5}
\begin{center}
  {\bf \Large Supplementary Material}
\end{center}
This supplementary material contains a review on the recent developments in multivariate normality (MVN) tests for i.i.d. data based on methods other than the skewness and kurtosis approaches, the chi-squared type and BHEP-type tests that are described in the body of the manuscript.

\setcounter{section}{0}
\section{Other generalizations of univariate normality tests}
The testing procedures for MVN reviewed in the main manuscript are all extensions of univariate techniques. In this section, we present four other recent generalizations that cannot be classified into any of the groups described there. 

\cite{szekely2005new} proposed a class of 
multivariate goodness-of-fit tests based on Euclidean distance between sample elements, and applied the tests for assessing MVN. The goodness-of-fit test statistic is defined by 
\[
\delta_{n,p}=n\left(\frac{2}{n}\sum_{j=1}^{n}\mbox{E}(\|\bX_{j}-\bX\|)-\mbox{E}(\|\bX-\bX'\|)-\frac{1}{n^2}\sum_{j,k=1}^{n}\|\bX_j-\bX_k\| \right),
\]
where $\bX$ and $\bX'$ are i.i.d. random vectors from the null distribution. $\delta_{n,p}/n$ is actually a von-Mises-statistic, or simply $V$-statistic \citep{mises1947asymptotic, hoeffding1948class}: $\delta_{n,p}/n=\frac{1}{n^2}\sum_{j,k=1}^{n}h(\bX_j,\bX_k)$, with kernel
\[
h(\bx,\by)=\mbox{E}(\|\bx-\bX\|)+\mbox{E}(\|\by-\bX\|)-\mbox{E}(\|\bX-\bX'\|)-\|\bx-\by\|,\;\bx,\by\in\bbR^{p}.
\]
Since the kernel $h(\bx,\by)$ for $p=1$ is closely related to the Cram{\'e}r-von Mises distance (see their Equation (17)), this test is a multivariate version of a Cram{\'e}r-von Mises type test. If the null distribution is $\mathcal{N}_{p}(\bmu,\bSigma)$, the test statistic for MVN is given by
\[
\delta_{n,p}=n\left(\frac{2}{n}\sum_{j=1}^{n}\mbox{E}(\|\bY_{j}-\bZ\|)-\mbox{E}(\|\bZ-\bZ'\|)-\frac{1}{n^2}\sum_{j,k=1}^{n}\|\bY_j-\bY_k\| \right),
\]
where $\bY_{j}, j=1,\ldots,n$ are the scaled residuals, and $\bZ$ and $\bZ'$ are i.i.d. random vectors from $\mathcal{N}_{p}(\textbf{\textit{0}},\bI_{p})$. The explicit form of $\delta_{n,p}$ is given by their Equation (8).

\cite{villasenor2009generalization} proposed a generalization of Shapiro--Wilk's test for MVN. Suppose $\bZ=(Z_1,\ldots,Z_n)^{\top}$ is the set of order statistics from a standard normal random sample, and $\mbox{E}(\bZ)=\bom$ and $\mbox{cov}(\bZ)=\bV$. If a set of ordered sample, $\bX^{*}=(X_1,\ldots,X_n)^{\top}$, comes from a normal distribution $\mathcal{N}(\mu,\sigma^2)$, then on a normal probability plot, $X_{i}=\mu+\sigma Z_i, i=1,\ldots,n$. Hence, the best linear unbiased estimates of $\mu$ and $\sigma$ are the generalized least square estimates that minimize the quadratic form $(\bX^{*}-\mu \boldsymbol{1}-\sigma\bom)^{\top}\bV^{-1}(\bX^{*}-\mu \boldsymbol{1}-\sigma\bom)$, where $\boldsymbol{1}$ is the vector of ones of length $n$; that is, $\widehat{\mu}=\overbar{X}$ and $\widehat{\sigma}=(\bom^{\top}\bV^{-1}\bom)^{-1}\bom^{\top}\bV^{-1}\bX^{*}$. Let $s^2=\frac{1}{n-1}\sum_{i=1}^{n}(X_i-\overbar{X})^2$ be the unbiased estimate of $\sigma^2$. The Shapiro-Wilk test statistic is defined by
\begin{equation}
	W=\frac{R^{4}\widehat{\sigma}^2}{(n-1)C^{2}s^2}=\frac{b^2}{(n-1)s^2}=\frac{(\ba^{\top}\bX^{*})^2}{(n-1)s^2}=\frac{\left(\sum_{i=1}^{n}a_{i}X_{i}\right)^2}{\sum_{i=1}^{n}(X_i-\overbar{X})^2},
	\label{eq:SW}
\end{equation}
where $R^2=\bom^{\top}\bV^{-1}\bom$, $C^2=\bom^{\top}\bV^{-1}\bV^{-1}\bom$, $\ba^{\top}=(\bom^{\top}\bV^{-1}\bV^{-1}\bom)^{-1/2}\bom^{\top}\bV^{-1}$ and $b=R^{2}\widehat{\sigma}/C$. $W$ is close to one under normality. For the multivariate random sample $\bX_1,\ldots,\bX_n$, using the fact that under $H_0$, the scaled residuals $\bY_j, j=1,\ldots,n$ have a distribution close to $\mathcal{N}_{p}(\textbf{\textit{0}},\bI_{p})$, which means that the coordinates of $\bY_j$, denoted by $Y_{1j},\ldots,Y_{pj}$, are approximately i.i.d. random variables from $\mathcal{N}(0,1)$, \cite{villasenor2009generalization} proposed a test statistic for assessing MVN:
\[
W^{*}=\frac{1}{p}\sum_{i=1}^{p}W_{Y_i},
\]
where $W_{Y_i}$ is the Shapiro--Wilk's test statistic evaluated on the $i$th coordinate of the transformed observations, $Y_{i1},\ldots,Y_{in}$, $i=1,\ldots,p$. 

\cite{majerski2010approximations} derived some approximations to the most powerful invariant (MPI) tests for MVN. Exact MPI tests for univariate normality have been well studied for some specific alternatives, such as uniform, double exponential, exponential, and Cauchy \citep{uthoff1970optimum, uthoff1973most, franck1981most}. Exact, but computationally cumbersome, MPI tests for binomality have been developed by \cite{szkutnik1988most} for two specific alternatives only, i.e., bivariate uniform and bivariate exponential, and MPI tests for $p>2$ have not been studied so far. \cite{majerski2010approximations} constructed approximations to the tests presented by \cite{szkutnik1988most} using the Laplace expansion for integrals, and showed that the approximations are asymptotically equivalent to the likelihood ratio (LR) tests, as is the case in the univariate setting. Furthermore, the authors extended their results to the cases of $p>2$, which are, however, limited to low-dimensional cases, due to the computational accuracy and complexity of numerical integration approximations for high dimensions. By showing in simulation studies that the MPI tests have practically the same powers as the LR tests, they provided a strong motivation for using the simple and fast LR test procedures for higher dimensions.

\cite{kim2018likelihood} presented extensions of the univariate omnibus LR tests, which are based on empirical distribution functions (EDF), to the tests for MVN. \cite{zhang2002powerful} proposed a goodness-of-fit LR test statistic based on the univariate observations $X_1,\ldots,X_n$:
\[
Z_{A}=-\sum_{i=1}^{n}\left[ \frac{\log F_{0}(X_{(i)})}{n-i+1/2} + \frac{\log \{1-F_{0}(X_{(i)})\}}{i-1/2} \right],
\]
where $F_{0}(\cdot)$ is the null distribution function, and $X_{(i)}$'s are the order statistics. For the multivariate sample $\bX_1,\ldots,\bX_n$, the scaled residuals $\bY_j, j=1,\ldots,n,$ are approximately $\mathcal{N}_p(\textbf{\textit{0}},\bI_{p})$, which indicates that the coordinates of $\bY_j$, denoted by $Y_{1j},\ldots,Y_{pj}$, and furthermore, all the elements $Y_{ij}$, $i=1,\ldots,p$, $j=1,\ldots,n$, are approximately i.i.d. random variables from $\mathcal{N}(0,1)$. \cite{kim2018likelihood} thus suggested the test statistic using the coordinate-wise characterization as
\[
Z_{A}^{*}=\frac{1}{p}\sum_{i=1}^{p}Z_{A}^{(i)}=-\frac{1}{p}\sum_{i=1}^{p}\sum_{j=1}^{n}\left[ \frac{\log \Phi(Y_{i(j)})}{n-i+1/2} + \frac{\log \{1-\Phi(Y_{i(j)})\}}{i-1/2} \right],
\]
where $Z_{A}^{(i)}$, $i=1,\ldots,p$, is the univariate LR statistic for the $i$th component, $Y_{i(j)}$ is the $j$th order statistic of $Y_{i1},\ldots,Y_{in}$, and $\Phi(\cdot)$ is the cumulative distribution function (CDF) of $\mathcal{N}(0,1)$. The second test statistic based on the element-wise characterization is given by
\[
Z_{A}^{**}=-\sum_{i=1}^{m}\left[ \frac{\log  \Phi(Y_{(i)})}{m-i+1/2} + \frac{\log \{1- \Phi(Y_{(i)})\}}{i-1/2} \right],
\]
where $Y_{(i)}$ is the $i$th order statistic of $\mbox{vec}\{(\bY_1,\ldots,\bY_n)^{\top}\}$, and $m=np$.

\section{Multiple test procedures}
In this section, we present two recent testing procedures that combine multiple tests for MVN. 

\cite{tenreiro2011affine} proposed a multiple test procedure that combines a finite set of affine invariant test statistics for MVN through an improved Bonferroni method. The test statistic is 
\begin{equation}
	T_{n}(u)=\max_{h\in H}\{T_{h}-c_{n,h}(u)\},
	\label{eq:tnu}
\end{equation}
where $u\in [0,1]$, $T_{h}, h\in H$, is any finite family of affine invariant test statistics for MVN, and $c_{n,h}(u)$ is the quantile of order $1-u$ of $T_{h}$ under $H_0$.  For a significance level $\alpha\in [0,1]$, the multiple test procedure rejects the null hypothesis of MVN whenever $T_{n}(u_{n,\alpha})>0$, where $u_{n,\alpha}=\sup \{u\in [0,1]: P_{\phi}\{T_{n}(u)\}>0\leq \alpha  \}$, and $\phi$ is the density for $\mathcal{N}_p(\textbf{\textit{0}},\bI_{p})$.
The usefulness of such an approach is illustrated by a multiple test combining some of the most recommended tests, i.e., the Mardia's skewness and kurtosis tests \citep{mardia1970measures, mardia1974applications} (the former performs well for skewed or long-tailed alternatives, and the latter for short-tailed alternatives), and the BHEP tests with two choices of the tuning parameter $\beta$ in the statistic $T_{n,\beta}$:
\begin{equation}
	\beta_{S}=0.448+0.026p \;\;\mbox{   and   }\;\; \beta_{L}=0.928+0.049p,
	\label{eq:betaLS}
\end{equation}  
which depend on the dimension $p$ for $2\leq p \leq 15$, and are identified from simulation studies by \cite{tenreiro2009choice} based on their distinct behavior patterns for the empirical power of BHEP tests as a function of $\beta$. $\beta_S$ is shown to be suitable for short-tailed or high-moment alternatives, while $\beta_L$ is appropriate for long-tailed or moderately skewed alternative distributions. The multiple test procedure was further studied in \cite{tenreiro2017new}, who combined BHEP tests with four different values of $\beta$ in the statistic $T_{n,\beta}$: two non-extreme choices, where $\beta=\beta_S$ and $\beta=\beta_L$ as defined in Equation~(\ref{eq:betaLS}), and two extreme cases, where $\beta\rightarrow 0$ and $\beta\rightarrow \infty$. 

\cite{zhou2014powerful} proposed a test that combines the univariate Shapiro-Wilk test for projected data and Mardia's multivariate kurtosis test. The Shapiro-Wilk test statistic $W$, given by Equation~(\ref{eq:SW}), can be used to detect non-normality in univariate projections of the scaled residuals $\bY_j, j=1,\ldots,n$, in the direction $\btheta$:
\[
G_{n}(\btheta)=W(\btheta^{\top}\bY_1,\ldots,\btheta^{\top}\bY_n).
\]
While \cite{fattorini1986remarks} considered a test for detecting non-normality of multivariate data projected in the most ``extreme'' direction among $\|\bY_j\|^{-1}\bY_j$, $j=1,\ldots,n$, corresponding to the smallest $G_n$ value, \cite{zhou2014powerful} considered the $p$ most ``extreme'' directions corresponding to the $p$ smallest $G_n$ values evaluated at the same random directions, denoted by $\Theta_{1}$. They also consider the $p$ unit vector directions $\boldsymbol{e}_{j}=(0,\ldots,0,1,0,\ldots,0)^\top$, $j=1,\ldots,p$, denoted by $\Theta_2$, which project the multivariate data to the $p$ marginal variates that are also normal under $H_0$. The new test statistic, incorperating Mardia's multivariate kurtosis statistic, is defined as
\[
T_{n,c}=1-\frac{1}{2p}\sum_{\btheta\in\Theta_{1}\cup\Theta_{2}}G_{n}(\btheta)I_{A}
\] 
where $A=\{c_1\leq MK \leq c_2\}$ with $c_1$ and $c_2$ being certain percentiles (e.g., $1\%$ and $99\%$, respectively) of the MK statistic given by Equation (1) in the body of this article, and $I_A$ is the indicator function with a value of $1$ if $A$ is true and $0$ otherwise. 

\baselineskip=20pt

\end{document}